\documentclass[aps,superscriptaddress,eqsecnum,nofootinbib,preprintnumbers]{revtex4}
\usepackage{graphicx,epsfig}
\usepackage{amsmath}
\usepackage{amssymb}
\usepackage{color}
\usepackage{subfig}
\usepackage{multirow}

\newcommand{\be}{\begin{eqnarray}}
\newcommand{\ee}{\end{eqnarray}}
\newcommand{\bea}{\begin{eqnarray}}
\newcommand{\nn}{\nonumber}
\newcommand{\eea}{\end{eqnarray}}

\newcommand{\PR}[1]{\ensuremath{\left[#1\right]}} % parenteses rectos do tamanho adequado
\newcommand{\PC}[1]{\ensuremath{\left(#1\right)}} % parenteses curvos do tamanho adequado
\newcommand*{\LargerCdot}{\raisebox{-0.25ex}{\scalebox{0.9}{$\cdot$}}}

\begin{document}

\title{Dynamics of cosmological perturbations in modified Brans-Dicke cosmology
with matter-scalar field interaction}

\author{Georgios Kofinas}
\email{gkofinas@aegean.gr} \affiliation{Research Group of
Geometry, Dynamical Systems and Cosmology,
Department of Information and Communication Systems Engineering\\
University of the Aegean, Karlovassi 83200, Samos, Greece}

\author{Nelson A. Lima}
\email{n.lima@thphys.uni-heidelberg.de}
%\homepage[]{Your web page}
%\thanks{}
%\altaffiliation{}
\affiliation{Institut f\"{u}r Theoretische Physik, Ruprecht-Karls-Universit\"{a}t Heidelberg, \\ Philosophenweg 16, 69120 Heidelberg, Germany}
%\author{Luca Amendola}
%\email{l.amendola@thphys.uni-heidelberg.de}
%\affiliation{Institut f\"{u}r Theoretische Physik, Ruprecht-Karls-Universit\"{a}t Heidelberg, \\ Philosophenweg 16, 69120 Heidelberg, Germany}

\begin{abstract}

In this work we focus on a novel completion of the well-known Brans-Dicke theory that
introduces an interaction between the dark energy and dark matter sectors, known as complete
Brans-Dicke (CBD) theory. We obtain viable cosmological accelerating solutions that fit
Supernovae observations with great precision without any scalar potential $V(\phi)$. We use these
solutions to explore the impact of the CBD theory on the large scale structure by studying the dynamics
of its linear perturbations. We observe a growing behavior of the lensing potential $\Phi_{+}$ at
late-times, while the growth rate is actually suppressed relatively to $\Lambda$CDM, which allows
the CBD theory to provide a competitive fit to current RSD measurements of $f\sigma_{8}$. However,
we also observe that the theory exhibits a pathological change of sign in the effective
gravitational constant concerning the perturbations on sub-horizon scales that could pose a challenge
to its validity.
%Unable to be recast as a minimally coupled scalar-tensor theory, the CBD theory produces interesting phenomenology that could be used to constrain it
\end{abstract}

\maketitle

\section{Introduction \label{intro}}

Two decades after the discovery of the late-times accelerated expansion of our
Universe~\cite{accel1,accel2}, comprehending the physical nature behind the effect stands as one
of the more important challenges in modern physics.
In the standard model of cosmology, the $\Lambda$ cold dark matter ($\Lambda$CDM) model, a
negative-pressure cosmological constant $\Lambda$ makes the majority of the energy density in
the present cosmos and accelerates its expansion within the framework of Einstein's general
relativity (GR). While $\Lambda$ may be attributed to a vacuum energy, its observed value is
inexplicably small to theory (for a review on $\Lambda$ see \cite{lambdareview}).

Hence, modified theories of gravity (MGT) were introduced to explain our Universe's accelerated
expansion as an alternative to $\Lambda$CDM.
Scalar-tensor gravity theories are widely studied as alternatives to general relativity and can
play a significant role in the description of the early or late-times cosmic evolution.
However, presently, it is not clear that scalar-tensor
theories, such as Brans-Dicke (BD) \cite{bd}, Galileon theory \cite{Gall}, $f(R)$ models
\cite{frreview}, and many others embedded within the Horndeski formalism \cite{horndeski:74} can
provide self-accelerating solutions compatible with cosmological observations \cite{lombriser:16},
and hence be genuine alternatives to $\Lambda$ or dark energy (DE) (for a review on MGT and DE see
\cite{reviewall,joyce:16,nojiri:2010wj}).

In BD gravity in particular, the scalar field forms the dark energy or can play a role
in the early Universe history, but also controls the evolution of the gravitational constant.
However, it is well known that in standard
BD theory self-accelerating solutions are not compatible with Solar-system constraints
\cite{ssbd,ssbd2} or even the latest cosmic microwave background (CMB) results
\cite{planckbd,planckbd2}, as these require a negative, order-unity Brans-Dicke parameter
$\omega_{\rm{BD}}$ \cite{bd1,bd2}. In order to avoid this issue one either adds a self-interacting
potential \cite{bd3,bd4,bd7}, considers a field or time-dependent $\omega_{\rm{BD}}$ \cite{bd5},
but even then the problem is not completely solved. Additionally, non-minimal couplings to matter
have been considered in Refs.~\cite{bd8,bd9,bd10,bd11}.

Most of the cosmological models consider that the evolution of dark matter and dark energy occur
separately. This means that the matter Lagrangian is added minimally to the action. In Ref.~\cite{Bertolami:2007zm} it was argued that there are observational evidences
which indicate a dark matter-dark energy interaction and violation of the equivalence principle
between baryons and dark matter. There is a raising activity in cosmology in the study of such
interacting models (e.g. \cite{Costa:2013sva,delCampo:2015vha}) which can also help to the
solution of the coincidence problem \cite{Amendola:1999qq,Zimdahl:2003wg}. Usually, such interactions
are chosen arbitrarily and do not arise by any physical theory. In the context of BD gravity an
energy exchange model with a modified wave equation for the scalar field was considered in Ref.~\cite{Clifton:2006vm} (for other approaches with modified equations of motion see
\cite{Smalley:1974gn,Bertolami:1999dp,Das:2008iq,Chakraborty:2008xt}).

In this work, we focus on a novel extension of the BD theory introduced in Ref.~\cite{Kofinas:2015nwa},
where the simple wave equation of the scalar field was preserved, while the standard conservation
of matter was relaxed. Analyzing exhaustively the Bianchi identities, three completions of
Brans-Dicke gravity were found to be the only theories which are unambiguously determined from
consistency. Here, we will focus on the first of these theories that we will call for brevity
as complete Brans-Dicke theory (CBD). This theory has an extra parameter $\nu$ that naturally
appears as an integration constant, which controls the energy exchange between the dark energy
and matter sectors, and when set to zero allows one to recover the standard Brans-Dicke
field equations. Although BD gravity was initially formulated in terms of an action solely based
on dimensional arguments with the matter Lagrangian being minimally coupled, CBD theory was derived
at the level of the equations of motion. The reason is that in the presence of interactions between
the matter Lagrangian and the scalar field, there is an infinite number of actions that can be constructed which recover the standard BD action in the absence of interactions.

A discussion on the action of CBD
theory was given in Ref.~\cite{Kofinas:2015sjz}, where it was shown that, for a matter
Lagrangian that vanishes on-shell (such as pressureless dust, for example), the theory can not be
recast as a minimally coupled scalar-tensor theory in either the Einstein or Jordan frame. Hence,
it should be able to produce interesting phenomenology that cannot be associated to standard
Brans-Dicke gravity. Furthermore, and more importantly, the complete BD theory is capable of
providing self-accelerating solutions for negative values of this new constant in the absence of
a scalar field potential \cite{Kofinas:2016fcp}. However, these solutions have not yet been fully
explored.

Therefore, in this work, we set out to study the impact these solutions can have on the large scale
structure of the Universe by analyzing the dynamics of their linear perturbations. Presently, there
is an effort to obtain constraints on modified theories of gravity on larger scales that are
competitive with those we have on Solar-system scales, with a surge of surveys in the next decade
that will improve our knowledge of the Universe on cosmological scales, such as the Dark Energy
Survey (DES) \cite{des}, the extended Baryon Oscillation Spectroscopic Survey (eBOSS) \cite{eboss}
and the Euclid survey \cite{euclid} (for a review on cosmological tests of gravity see
\cite{koyama:15}). Hence, it is of paramount importance to understand how a particular theory
modifies the observable Universe.

This paper is organized as follows: in Sec.~\ref{seccosmo} we introduce the complete Brans-Dicke
theory and its field equations, and also extend its background solutions presented in
Ref.~\cite{Kofinas:2016fcp} to high-redshifts. Then, in Sec.~\ref{perttheory} we derive the
full set of perturbed equations of motion and present them in the Newtonian and synchronous gauges.
In Sec.~\ref{lensingpot} we present the dynamical first-order differential equations for the
lensing potential, $\Phi_{+}$, and the slip between the Newtonian potentials, $\chi$, that we
numerically evolve to study the dynamics of the linear perturbations. We then derive the sub-horizon
approximation for the Newtonian potentials in Sec.~\ref{subhorsec}, and compute the evolution of the
growth rate $f\sigma_{8}$ in Sec.~\ref{sec_growth}, concluding in Sec.~\ref{conclusion}.

\section{Cosmology in the complete Brans-Dicke theory \label{seccosmo}}

We consider the complete Brans-Dicke theory presented in \cite{Kofinas:2015nwa} and
described by the following equations
\begin{eqnarray}
&&\!\!\!\!\!\!\!G^{\mu}_{\,\,\,\nu}\!=\!\frac{8\pi}{\phi}
(T^{\mu}_{\,\,\,\nu}+\mathcal{T}^{\mu}_{\,\,\,\,\nu})
\label{elp}\\
&&\!\!\!\!\!\!\!T^{\mu}_{\,\,\,\nu}\!=\!\frac{\phi}{2\lambda(\nu\!+\!8\pi\phi^{2})^{2}}\Big{\{}
2\big[(1\!+\!\lambda)\nu\!+\!4\pi(2\!-\!3\lambda)\phi^{2}\big]\phi^{;\mu}\phi_{;\nu}
\!-\!\big[(1\!+\!2\lambda)\nu\!+\!4\pi(2\!-\!3\lambda)\phi^{2}\big]\delta^{\mu}_{\,\,\,\nu}
\phi^{;\rho}\phi_{;\rho} \Big{\}}
\!+\!\frac{\phi^{2}}{\nu\!+\!8\pi\phi^{2}}
\big(\phi^{;\mu}_{\,\,\,\,;\nu}\!-\!\delta^{\mu}_{\,\,\,\nu}\Box\phi\big)
\nn\\
\label{utd}\\
&&\!\!\!\!\!\!\!\Box\phi\!=\!4\pi\lambda\mathcal{T}\label{lrs}\\
&&\!\!\!\!\!\!\!\mathcal{T}^{\mu}_{\,\,\,\,\nu;\mu}\!=\!\frac{\nu}{\phi(\nu\!+\!8\pi\phi^{2})}
\mathcal{T}^{\mu}_{\,\,\,\,\nu}\phi_{;\mu}\,.
\label{idj}
\end{eqnarray}
Compared to the standard Brans-Dicke theory, the new characteristic of these equations is the
appearance of the parameter $\nu$, with dimensions mass to the fourth, which enters the
gravitational field equations. And, at the same time, it violates the exact conservation
of the matter energy-momentum tensor $\mathcal{T}^{\mu}_{\,\,\,\,\nu}$ in Eq.~(\ref{idj}).
The parameter $\lambda\neq 0$ is related to the standard Brans-Dicke parameter $\omega_{\rm{BD}}=
\frac{2-3\lambda}{2\lambda}$. The system (\ref{elp})-(\ref{idj}) reduces for $\nu=0$ to the
Brans-Dicke equations of motion (in units where the velocity of light is set to unit)
\begin{eqnarray}
G^{\mu}_{\,\,\,\nu}\!\!&=&\!\!\frac{8\pi}{\phi}(T^{\mu}_{\,\,\,\nu}+\mathcal{T}^{\mu}_{\,\,\,\,\nu})
\label{kns}\\
T^{\mu}_{\,\,\,\nu}\!\!&=&\!\!\frac{2-3\lambda}{16\pi\lambda\phi}\Big(
\phi^{;\mu}\phi_{;\nu}\!-\!\frac{1}{2}\delta^{\mu}_{\,\,\,\nu}
\phi^{;\rho}\phi_{;\rho} \Big)
\!+\!\frac{1}{8\pi}\big(\phi^{;\mu}_{\,\,\,\,;\nu}\!-\!\delta^{\mu}_{\,\,\,\nu}\Box\phi\big)
\label{qqd}\\
\Box\phi\!\!&=&\!\!4\pi\lambda\mathcal{T}\label{lwn}\\
\,\,\mathcal{T}^{\mu}_{\,\,\,\,\nu;\mu}\!\!&=&\!\!0\,
\label{jrk}
\end{eqnarray}
which is described by the action
\begin{equation}
S_{BD}=\frac{1}{16\pi}\int \!d^{4}x \,\sqrt{-g} \,\Big(\phi R-\frac{\omega_{BD}}{\phi}
g^{\mu\nu}\phi_{,\mu}\phi_{,\nu}\Big)+\int \!d^{4}x \,\sqrt{-g} \,L_{m}\,,
\label{jsw}
\end{equation}
where $L_{m}(g_{\kappa\lambda},\Psi)$ is the matter Lagrangian
depending on some extra fields $\Psi$. The system of equations (\ref{elp}), (\ref{idj}) will be
analyzed for both a cosmological background and for its perturbations.

For the theory (\ref{elp})-(\ref{idj}), a statistically spatially homogeneous and isotropic universe
with Friedmann-Robertson-Walker (FRW) metric has been studied in \cite{Kofinas:2016fcp}. Here, we
consider the spatially flat case with background metric
\begin{equation}
d\bar{s}^2 = - a^2d\tau^2 + a^{2}(\tau)\delta_{ij}dx^{i}dx^{j}~,
\label{metric}
\end{equation}
where $\tau$ is the conformal time and we will denote with an overdot the derivative with respect
to $\tau$. The modified Friedmann equations, the dynamical Brans-Dicke scalar field equation and
the energy-momentum conservation equation are given by
\begin{eqnarray}
&&\mathcal{H}^{2}=\frac{8\pi}{3\varphi}\rho a^{2}-\frac{8\pi\varphi}{\nu\!+\!8\pi\varphi^{2}}
\mathcal{H}\dot{\varphi}+\frac{4\pi}{3\lambda}\,\frac{\nu\!+\!4\pi(2\!-\!3\lambda)\varphi^{2}}
{(\nu\!+\!8\pi\varphi^{2})^{2}}\dot{\varphi}^{2}
\label{eji}\\
&&2\dot{\mathcal{H}}+\mathcal{H}^{2}=-\frac{8\pi}{\varphi}\Big[
pa^{2}+\frac{\varphi}{2\lambda}\,\frac{(1\!+\!2\lambda)\nu\!+\!4\pi(2\!-\!3\lambda)\varphi^{2}}
{(\nu\!+\!8\pi\varphi^{2})^{2}}
\dot{\varphi}^{2}+\frac{\varphi^{2}}{\nu\!+\!8\pi\varphi^{2}}
(\mathcal{H}\dot{\varphi}+\ddot{\varphi})\Big]
\label{ket}\\
&&\ddot{\varphi}+2\mathcal{H}\dot{\varphi}+4\pi\lambda(3p\!-\!\rho)a^{2}=0
\label{iet}\\
&&\dot{\rho}+3\mathcal{H}(\rho\!+\!p)=\frac{\nu}{\varphi(\nu\!+\!8\pi\varphi^{2})}
\rho\,\dot{\varphi}~,
\label{euf}
\end{eqnarray}
with $\mathcal{H}\equiv\frac{\dot{a}}{a}=aH$ the conformal Hubble factor, where
$H=\frac{1}{a}\frac{da}{dt}$ is the Hubble parameter ($dt=ad\tau$). The background scalar
field is denoted by $\varphi(\tau)$, while in the next section where perturbations will be
introduced, the total perturbed field will be $\phi=\varphi+\delta\phi$, with $\delta\phi$ representing
the perturbation. Equation (\ref{euf}) can be integrated into a simple expression for the evolution of
the matter energy density as a function of time
\begin{equation}
\rho=\frac{\rho_{\ast}}{a^{3(1+w)}}\frac{\varphi}{\sqrt{|\nu\!+\!8\pi\varphi^{2}}|}~,
\label{keg}
\end{equation}
where $\rho_{\ast}>0$ is an integration constant and it is assumed that $\varphi>0$.

We can write the Friedmann equations (\ref{eji}), (\ref{ket}) in a more familiar form
\begin{eqnarray}
&&\mathcal{H}^{2}=\frac{8\pi a^{2}}{3\varphi}\left(\rho+\rho_{\rm{DE}}\right)
\label{FR1}\\
&&2\dot{\mathcal{H}}+\mathcal{H}^{2}=-\frac{8\pi a^{2}}{\varphi}\left(p+p_{\rm{DE}}\right)\,,
\label{FR2}
\end{eqnarray}
where we have defined the effective dark energy and effective dark
pressure as
\begin{eqnarray}
&&\rho_{\rm{DE}}a^{2}\equiv -\frac{3\varphi^2}{\nu\!+\!8\pi\varphi^{2}} \mathcal{H}\dot{\varphi}
+\frac{\varphi}{2\lambda}\,\frac{\nu\!+\!4\pi(2\!-\!3\lambda)\varphi^{2}}
{(\nu\!+\!8\pi\varphi^{2})^{2}}\dot{\varphi}^{2}
\label{rhoDE}\\
&&p_{\rm{DE}}a^{2}\equiv
\frac{\varphi}{2\lambda}\,\frac{(1\!+\!2\lambda)\nu\!+\!4\pi(2\!-\!3\lambda)\varphi^{2}}
{(\nu\!+\!8\pi\varphi^{2})^{2}}
\dot{\varphi}^{2}+\frac{\varphi^{2}}{\nu\!+\!8\pi\varphi^{2}}(\mathcal{H}\dot{\varphi}+\ddot{\varphi})~.
\label{pDE}
\end{eqnarray}
Then, according to (\ref{FR1}), the density parameters are defined as
\begin{equation}
\Omega_{\rm{m}}=\frac{8\pi\rho a^{2}}{3\varphi
\mathcal{H}^2}~,\,\,\,\,\,\,\,\,\,\Omega_{\rm{DE}}=\frac{8\pi\rho_{\rm{DE}}a^{2}}{3\varphi
\mathcal{H}^2}~. \label{qed}
\end{equation}

In Ref.~\cite{Kofinas:2016fcp}, the numerical background solutions were obtained integrating the
Friedmann and the scalar field equations backwards in time, from a present-day value of the scale
factor normalized to $1$, i.e. $a_{0}=1$.
Hence, the value of the integration constant $\rho_{\ast}$ was set so
that $\Omega_{\rm{m}}$ today, $\Omega_{\rm{m}}^{0}$, would be equal to a fixed value close to $0.30$.
Then, the units were chosen so that the initial value of the scalar field, $\varphi_{0}$,
was fixed to be $1$. The present-day value of the scalar field velocity $\dot{\varphi}_{0}$
and the parameters $\lambda,\nu$ were constrained so that
$\Omega_{\rm{DE}}^{0}$ has the value $1-\Omega_{\rm{m}}^{0}$ and also that the
value of the effective dark energy equation of state $w_{\rm{DE}}=p_{\rm{DE}}/\rho_{\rm{DE}}$
was close to $-1$ today, with matter domination at earlier times. Using this ``backward''
method, the solutions obtained provided self-acceleration at the present for different values of
$\nu$ and $\lambda$. However, the stability of the solutions obtained
with this method toward very high-redshifts is not guaranteed, which we have numerically checked.

In this work, we are interested in obtaining the evolution of linear perturbations from deep within
matter domination.  We attempt to perform a forward numerical evolution from a high-redshift
$z_{\rm{i}} \gg 1$, so the initial conditions are set at $z_{\rm{i}}=1000$.
We choose to use the logarithmic variable $N=\ln{a}$ as the integration variable, thus its initial
value is $N_{\rm{i}}=-6.91$ (while today we still have $a_{0}=1$). The system of equations (\ref{eji}),
(\ref{ket}), (\ref{iet}), after using equation (\ref{keg}), is written equivalently as
\begin{eqnarray}
&&\frac{4\pi}{3\lambda}\,\frac{\nu\!+\!4\pi(2\!-\!3\lambda)\varphi^{2}}
{(\nu\!+\!8\pi\varphi^{2})^{2}}{\varphi'}^{2}
-\frac{8\pi\varphi}{\nu\!+\!8\pi\varphi^{2}} \varphi'
+\frac{8\pi\rho_{\ast}e^{-N}}{3\mathcal{H}^{2}\sqrt{|\nu\!+\!8\pi\varphi^{2}|}}-1=0
\label{ofse}\\
&&\frac{2}{\mathcal{H}}\mathcal{H}'+\frac{4\pi}{\lambda}
\frac{(1\!+\!2\lambda)\nu\!+\!4\pi(2\!-\!3\lambda)\varphi^{2}}{(\nu\!+\!8\pi\varphi^{2})^{2}}
{\varphi'}^{2}
+\frac{8\pi\varphi}{\nu\!+\!8\pi\varphi^{2}}\Big(\frac{4\pi\lambda\rho_{\ast}\varphi e^{-N}}
{\mathcal{H}^{2}\sqrt{|\nu\!+\!8\pi\varphi^{2}|}}\!-\!\varphi'\Big)+1=0
\label{wrer}\\
&&\varphi''+\Big(2\!+\!\frac{\mathcal{H}'}{\mathcal{H}}\Big)\varphi'-
\frac{4\pi\lambda\rho_{\ast}\varphi e^{-N}}{\mathcal{H}^{2}\sqrt{|\nu\!+\!8\pi\varphi^{2}|}}
=0\,,\label{yhen}
\end{eqnarray}
where a prime denotes a derivative with respect to $N$. The system (\ref{ofse})-(\ref{yhen})
contains the integration constant $\rho_{\ast}$ and the parameters $\lambda,\nu$ that have to be
specified. It is a consistent system since equation (\ref{ofse}) is the constraint. The analysis of
this system can be made in two ways. In the first one, the quantity $e^{-N}\mathcal{H}^{-2}$ is
replaced from (\ref{ofse}) into (\ref{wrer}), (\ref{yhen}), and then, an autonomous second-order
differential equation for $\varphi$ arises. When this equation is solved for $\varphi(N)$, then
$\mathcal{H}(N)$ is found algebraically from (\ref{ofse}). In this method we need at the initial
time $N_{\rm{i}}$ the two initial conditions $\varphi_{\rm{i}}, \varphi'_{\rm{i}}$. In the second
way, equations (\ref{ofse}), (\ref{wrer}) are viewed as a system of two first-order differential
equations for $\varphi, \mathcal{H}$. Now, we need at the initial
time $N_{\rm{i}}$ the two initial values $\varphi_{\rm{i}}, \mathcal{H}_{\rm{i}}$ (of course,
$\varphi'_{\rm{i}}$ can be found from (\ref{ofse})).

From the physical point of view the evolution should be such that at early times the contribution
of the effective dark energy density is negligible, i.e. $\Omega_{\rm{DE}} \ll 1$. As seen from
(\ref{rhoDE}), the simplest condition in order for this to be achieved is to choose
$|\varphi'_{\rm{i}}|\ll 1$, and the standard GR behaviour is recovered at early times.
This implies from equation (\ref{ofse}) the value of $\rho_{\ast}$ in terms of the initial
values $\varphi_{\rm{i}},H_{\rm{i}}$, i.e.
$\rho_{\ast}=(3/8\pi)a_{\rm{i}}^{3}H_{\rm{i}}^{2}\sqrt{|\nu\!+\!8\pi\varphi_{\rm{i}}^{2}|}$.
Therefore, there are not three independent integration constants, but only two.
In $\rm{\Lambda CDM}$ there are two integration constants, namely $H_{\rm{i}},\rho_{\ast}$,
while the condition of negligible initial dark energy is automatically satisfied, since at $z_{\rm{i}}$
the matter term is $10^{9}$ times larger than the cosmological constant term, therefore the two
initial data are set at present in agreement with the values $H_{0},\Omega_{\rm{m}}^{0}$.

Then, we fix the free parameters $\lambda$ and $\nu$. From (\ref{rhoDE}) we need to
set $\lambda$ such that $\lambda\gtrsim |\varphi^{\prime}_{\rm{i}}|$  in order to keep
$\Omega_{\rm{DE}} \ll 1$. In this work, we choose $\lambda = 1$. In Ref.~\cite{Kofinas:2016fcp}, it was
shown that the condition $\nu+8\pi\varphi^2<0$ is successful in order to have
accelerating solutions today. Although acceleration also appeared in some cases
where the above quantity is positive, here we will assume the negative sign and set $\nu$ to
a high negative value of $-100$. Therefore, a solution should be restricted to the branch with
$\varphi<\varphi_{\infty}=\sqrt{|\nu|/(8\pi)}$, otherwise poles would appear in the equations, e.g.
in Eq.~(\ref{keg}) for the energy density. One word about the units is needed at this
point. Since $\varphi^{-1}$ plays the role of varying gravitational constant $G$, the scalar field
$\varphi$ has dimensions of mass squared. Therefore, dimensionless quantities $\hat{\varphi},\hat{\nu}$
can be defined as $\hat{\varphi}=G_{\!N}\varphi$, $\hat{\nu}=G_{\!N}^{2}\nu$, where $G_{\!N}$ is
Newton's constant. Then, in all the previous equations we should replace $\varphi$ by $\hat{\varphi}$,
$\nu$ by $\hat{\nu}$, and all $\rho_{\ast},\,\rho,\,p,\,\rho_{\rm{DE}},\,p_{\rm{DE}}$ should be
multiplied by $G_{\!N}$. In this sense, in the numerical analysis, when we say that
$\nu$ is $-100$, we strictly mean that $\hat{\nu}$ is $-100$, while an order one value of $\varphi$
basically means of $\hat{\varphi}$. Moreover, it should be noted that the parameter $\nu$ can be
totally absorbed in the system (\ref{ofse})-(\ref{yhen}) when the rescaling
$\varphi\rightarrow \varphi/\sqrt{|\nu|}$, $\rho_{\ast}\rightarrow \rho_{\ast}/\sqrt{|\nu|}$ is
performed.

Since $|\varphi'_{\rm{i}}|\ll 1$, in the first period of evolution it is $\varphi\approx \varphi_{\rm{i}}$,
thus in equation (\ref{eji}) the derivatives of $\varphi$ can be omitted and we obtain
$H^{2}\approx a_{\rm{i}}^{3}H_{\rm{i}}^{2}a^{-3}$, which is the behaviour of Einstein gravity in
matter era. Instead of having the unknown dimensionfull initial value $H_{\rm{i}}$ in the above
expression of $H^{2}$, as well as in $\rho_{\ast}$, we prefer to normalize $H_{\rm{i}}$ to the
central value $\hat{H}_0=67.8\,\rm{km/s/Mpc}$ coming from the latest Planck data, and parametrize
$H_{\rm{i}}$ in terms of the dimensionless quantity $\hat{\Omega}_{\rm{m}}$ as follows:
$a_{\rm{i}}^{3}H_{\rm{i}}^{2}=\hat{H}_{0}^{2}\hat{\Omega}_{\rm{m}}$. Therefore, it is
$H^{2}\approx \hat{H}_{0}^{2}\hat{\Omega}_{\rm{m}}a^{-3}$ initially, and
$\rho_{\ast}=(3/8\pi)\hat{H}_{0}^{2}\hat{\Omega}_{\rm{m}}\sqrt{|\nu\!+\!8\pi\varphi_{\rm{i}}^{2}|}$.
The quantity $\hat{\Omega}_{\rm{m}}$ can be interpreted as a fictitious value of
the density parameter $\Omega_{\rm{m}}$ corresponding to the energy density $\rho_{\ast}$.

It is obvious that since the initial data are set at an early epoch, the evolution of the equations
does not assure that the evolved theoretical today values $H(a\!=\!1)$, $\Omega_{\rm{m}}(a\!=\!1)$
will coincide with the actual today values $H_{0},\Omega_{\rm{m}}^{0}$. Of course, the values
$H_{0},\Omega_{\rm{m}}^{0}$ are known from observations not precisely, but with a small uncertainty.
The value of $\Omega_{\rm{m}}^{0}$ is close to $0.30$ according to the most recent constraints
\cite{planckparams}. Therefore, $H(a\!=\!1)$, $\Omega_{\rm{m}}(a\!=\!1)$ should be close to the
values $\hat{H}_{0}$, 0.3 respectively, still within local observable bounds.
As a result, the two integration constants $\hat{\Omega}_{\rm{m}},\varphi_{\rm{i}}$,
which determine the whole evolution, cannot be chosen arbitrarily, but should provide consistent
values of $H(a\!=\!1)$, $\Omega_{\rm{m}}(a\!=\!1)$.
In the following we will succeed such an agreement between these theoretical
and observed values by fixing appropriately the initial conditions. Actually, we will be more
precise than that and provide a very good fit to low-redshift supernovae data. As for the density parameter, it arises that at all times it is
$\Omega_{\rm{m}}=\hat{H}_{0}^{2}\hat{\Omega}_{\rm{m}}\sqrt{|\nu\!+\!8\pi\varphi_{\rm{i}}^{2}|}
/(e^{3N}H^{2}\sqrt{|\nu\!+\!8\pi\varphi^{2}|})$. Initially,
$\Omega_{\rm{m}}^{\rm{i}}=1$, thus the condition for $\rho_{\ast}$ is found as above. Today,
\begin{equation}
\Omega_{\rm{m}}(a\!=\!1)=\hat{\Omega}_{\rm{m}}\sqrt{\frac{|\nu\!+\!8\pi\varphi_{\rm{i}}^{2}|}
{|\nu\!+\!8\pi\varphi_{0}^{2}|}}\frac{\hat{H}_{0}^{2}}{H^2(a\!=\!1)}\,,
\label{present_day_omegmam}
\end{equation}
where we denote by $\varphi_{0}$ the value of $\varphi(a=1)$ resulting from the numerical evolution,
since there is no observational constraint on the present value of the scalar field to distinguish
between $\varphi_{0}$ and $\varphi(a\!=\!1)$. Although $\varphi^{-1}$ is interpreted as the
varying gravitational constant $G$, this does not mean that $\varphi_{0}$ equals $G_{\!N}^{-1}$.
It is actually expected that $\varphi_{0}$ is of the order of $G_{\!N}^{-1}$, but the precise
numerical value is an issue of the initial conditions appropriate to explain the current state of
the Universe determined by $H_{0},\Omega_{\rm{m}}^{0}$. According to (\ref{present_day_omegmam}),
successful $\hat{\Omega}_{\rm{m}},\varphi_{\rm{i}}$ should provide that $H(a\!=\!1)$ is approximately
equal to $\hat{H}_{0}$ and also that $\varphi_{0}$ agrees with the value provided by
(\ref{present_day_omegmam}) with $\Omega_{\rm{m}}(a\!=\!1)$ close to 0.3. Thus, it is not an easy
task to find such $\hat{\Omega}_{\rm{m}},\varphi_{\rm{i}}$.

Since we have assumed that $\nu\!+\!8\pi\varphi^{2}<0$, it will be verified numerically that
the scalar field grows in time (it is actually expected that $G$ decreases with time),
thus $\varphi_{0}>\varphi_{\rm{i}}$. Since $H(a\!=\!1), \Omega_{\rm{m}}(a\!=\!1)$
must be close to $\hat{H}_{0},0.3$, it arises from (\ref{present_day_omegmam}) that
$\hat{\Omega}_{\rm{m}}<0.3$. This will implicate a larger separation between the cosmological
evolutions predicted by the complete Brans-Dicke theory and $\Lambda$CDM at early-times than at
late-times. We have implemented a Brent algorithm that searches for the right
$\varphi_{\rm{i}}$ in order to yield the desired $\Omega_{\rm{m}}(a\!=\!1)$ for a chosen
$\hat{\Omega}_{\rm{m}}$. The latter was fine-tuned to produce a value of $H(a=1)$ that is compatible
with current observations. In the figures shown in this section we have used
$\hat{\Omega}_{\rm{m}}=0.17$, $\varphi_{\rm{i}}=0.029$, thus $\varphi_{0}=1.773$ and
$H_{\rm{i}}=\hat{H}_{0}\sqrt{\hat{\Omega}_{\rm{m}}}\,(1+z_{\rm{i}})^{3/2}=13058 \hat{H}_{0}$.
The function $a(\tau)$ can be found numerically from the numerical solution $\mathcal{H}(N)$.

Another equivalent system of differential equations can be presented which eliminates the initial
condition $\hat{\Omega}_{\rm{m}}$ and at the same time it only needs to conform with a
consistent value of $\Omega_{\rm{m}}(a\!=\!1)$, thus it facilitates the search for appropriate initial
conditions. A rewriting of Eqs.~(\ref{ofse}), (\ref{wrer}) gives a system for the evolution of
$\varphi$ and $\Omega_{\rm{m}}$ as
\begin{eqnarray}
&&\frac{4\pi}{3\lambda}\,\frac{\nu\!+\!4\pi(2\!-\!3\lambda)\varphi^{2}}
{(\nu\!+\!8\pi\varphi^{2})^{2}}{\varphi'}^{2}
-\frac{8\pi\varphi}{\nu\!+\!8\pi\varphi^{2}} \varphi'
+\Omega_{\rm{m}}-1=0\label{kwree}\\
&&\frac{\Omega_{\rm{m}}'}{\Omega_{\rm{m}}}-\frac{4\pi}{\lambda}
\frac{(1\!+\!2\lambda)\nu\!+\!4\pi(2\!-\!3\lambda)\varphi^{2}}{(\nu\!+\!8\pi\varphi^{2})^{2}}
{\varphi'}^{2}
-\frac{4\pi\varphi}{\nu\!+\!8\pi\varphi^{2}}\big(3\lambda\varphi\Omega_{\rm{m}}
\!-\!4\varphi'\big)=0\,.\label{kjwr}
\end{eqnarray}
Initially it is $\Omega_{\rm{m}}^{\rm{i}}=1$ and only the initial condition $\varphi_{\rm{i}}$ is
free. Moreover, this system does not need to match the value $H(a\!=\!1)$, but only that of
$\Omega_{\rm{m}}(a\!=\!1)$. Therefore, scanning the parameter $\varphi_{\rm{i}}$ to provide
$\Omega_{\rm{m}}(a\!=\!1)=0.30$ is relatively easier, and the same values of
$\varphi_{\rm{i}}, \varphi_{0}$ are found as above. From the numerical solutions $\varphi(N),
\Omega_{\rm{m}}(N)$, a suitable $\hat{\Omega}_{\rm{m}}$ is selected as before that provides
algebraically the function $H(N)$ which possesses sufficient fitting to the supernovae.

\begin{figure}[t!]
\begin{minipage}{.5\linewidth}
\centering
\subfloat[]{\label{main:a}\includegraphics[scale=.40]{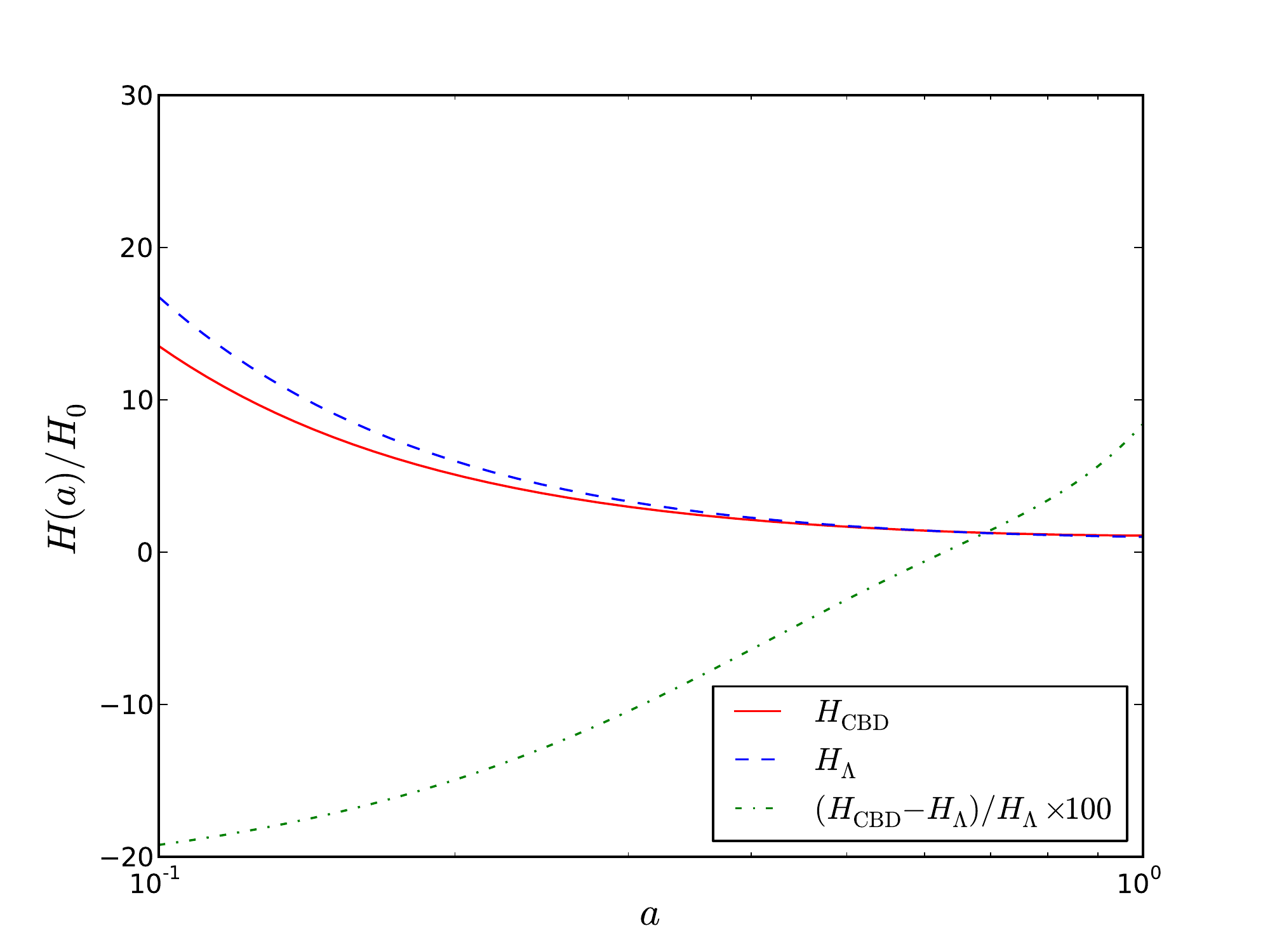}}
\end{minipage}%
\begin{minipage}{.5\linewidth}
\centering
\subfloat[]{\label{main:b}\includegraphics[scale=.40]{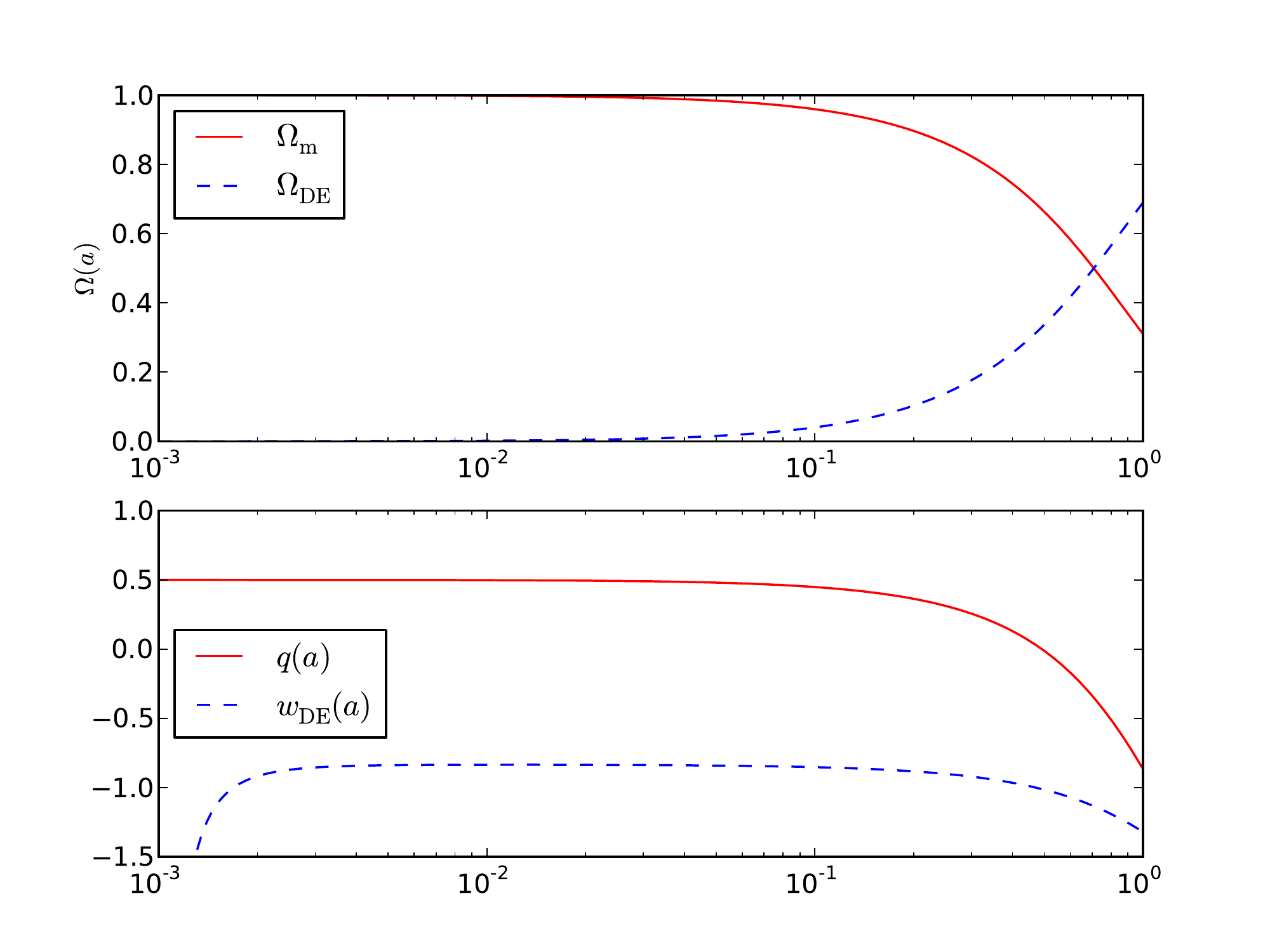}}
\end{minipage}\par\medskip
\begin{minipage}{.5\linewidth}
\centering
\end{minipage}
\caption{\label{back_evol}We plot the background evolution predicted by the complete Brans-Dicke
for the following choice of parameters: $\lambda = 1$, $\nu = -100$,
$\hat{\Omega}_{\rm{m}}=0.17$ and $\varphi_{\rm{i}}=0.029$. In (a)
we have the evolution of the Hubble parameter against $\Lambda$CDM; in (b) we plot the evolution
of $\Omega_{\rm{m}}$ and $\Omega_{\rm{DE}}$, together with the deceleration parameter $q$ and the
effective dark energy equation of state $w_{\rm{DE}}$.}
\end{figure}

In Fig.~\ref{back_evol} we plot the background evolution predicted by CBD according to the
explanations of the previous paragraphs. In Fig.~\ref{back_evol} (a)
we have the Hubble parameter as a function of the scale factor, compared against the evolution
predicted by the standard model, $\Lambda$CDM. As expected, we have a larger separation between
both cosmologies at earlier times. Today, we have a less than $10\%$ difference between both models, with the present-day value of $H$ predicted by the CBD equal to $H(a=1) = 73.4 \,\, \rm{km/s/Mpc}$, compatible with local measurements of the Hubble parameter \cite{loch1,loch2,loch3}.

Then, in Fig.~\ref{back_evol} (b) we plot $\Omega_{\rm{m}}(a)$ and $\Omega_{\rm{DE}}(a)$. We see that
our model provides a stable matter dominated phase that is gradually overtaken by the effective dark
energy component close to the present, yielding $\Omega_{\rm{m}}^0=0.30$ and
$\Omega_{\rm{DE}}^0=0.7$.
We also note that the flatness of Universe is guaranteed as we have numerically checked to have
$\Omega_{\rm{m}} + \Omega_{\rm{DE}} = 1$ throughout the cosmological evolution. The viability of
our model is further corroborated by the evolution of the
deceleration parameter $q = - 1 -\frac{1}{H^{2}}\frac{dH}{dt}=-\frac{\dot{\mathcal{H}}}
{\mathcal{H}^{2}}$ in Fig.~\ref{back_evol} (b), where we clearly observe the transition from
a decelerating to an accelerating dark energy dominated Universe close to the present-day.
We have also checked that the model asymptotically tends to the value $\varphi_{\infty}$ without
crossing it, therefore avoiding any singularity on the Hubble parameter and its derivatives.
As $\varphi$ approaches $\varphi_{\infty}$, the first term on the r.h.s. of Eq.~(\ref{rhoDE}),
which is positive, becomes enhanced and dominates the negative second term, ensuring a positive
$\Omega_{\rm{DE}}$.

\begin{figure}[t!]
\begin{center}
\includegraphics[scale = 0.50]{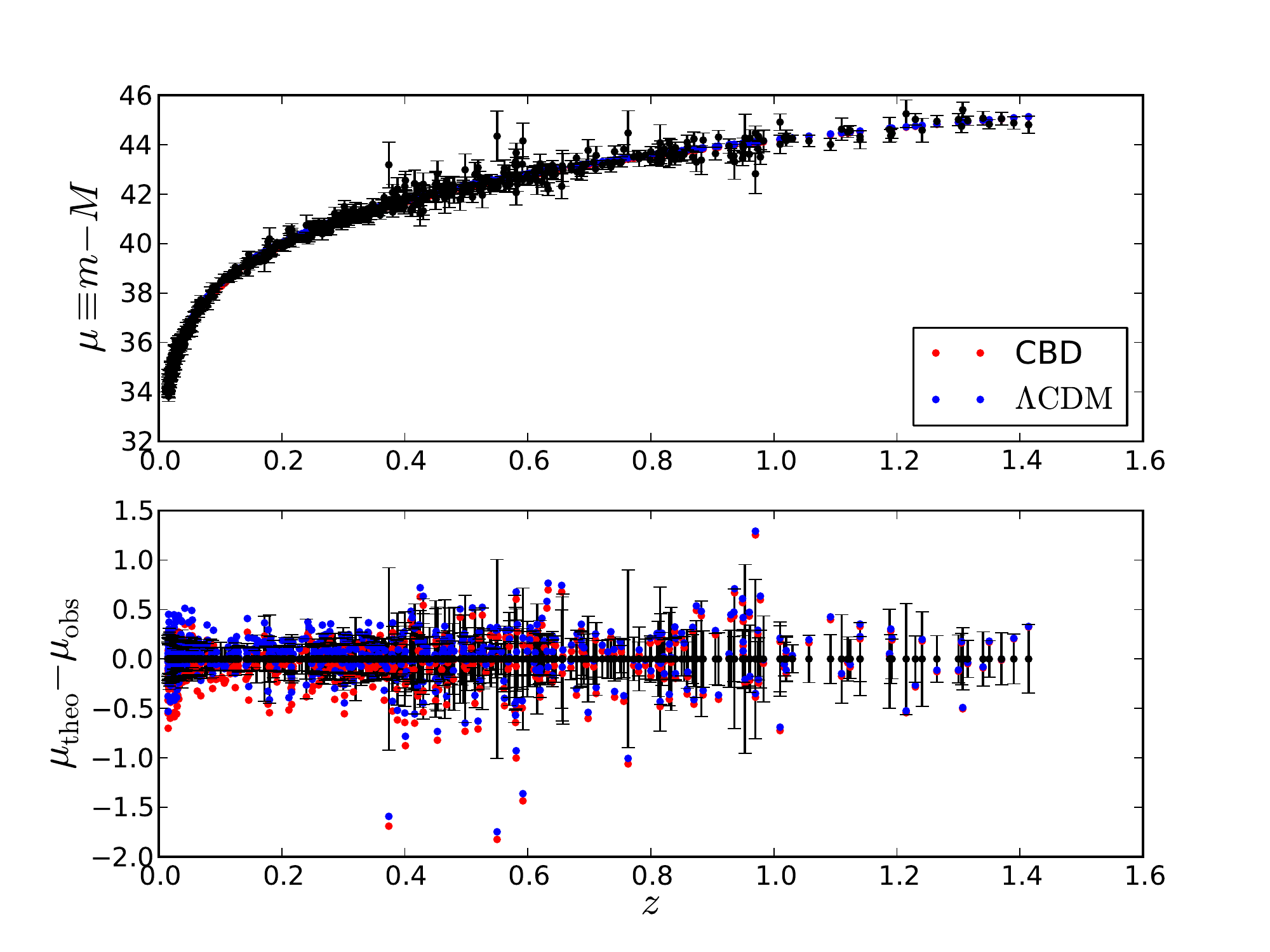}
\caption{\label{supernova_comp} We plot the distance moduli $\mu \equiv m - M$ predicted by
our model (using the parameters of Fig.~\ref{back_evol}) and $\Lambda$CDM, and compare to the
Union$2.1$ compilation from the Supernova Cosmology Project \cite{scp_supernova}. For the errorbars,
we adopted the covariance matrix without systematics.}
\end{center}
\end{figure}

Still in Fig.~\ref{back_evol} (b), we also have the evolution of the effective dark energy equation of state,
$w_{\rm{DE}}$, where we see that our solution predicts a phantom behavior today by having
$w_{\rm{DE},0} < -1$. For completeness, we also comment on the early-time behavior of $w_{\rm{DE}}$,
where we note that the effective equation of state tends to increasingly larger negative values.
This is a consequence of the way we have set our initial conditions. Eq.~(\ref{rhoDE}) shows that
$\rho_{\rm{DE}}$ will tend to negligibly smaller values at early-times the closer we set
$\varphi_{\rm{i}}^{\prime}$ to zero, leading to larger negative values in $w_{\rm{DE}}$.
This has, however, no discernible impact in the background evolution we predict, as
$\Omega_{\rm{DE}}$ is also negligible at the epoch we set the initial conditions.

Lastly, in Fig.~\ref{supernova_comp}, we compare our model to data from the Union$2.1$ compilation
of $580$ Type Ia supernova \cite{scp_supernova}, from which we adopt the covariance matrix without
the presence of systematics. We see that our model fits the data with remarkable precision,
comparable to $\Lambda$CDM, without the presence of a potential. Hence, having a complete
Brans-Dicke model that predicts a viable background history and fits existing data, we can
proceed to obtain the evolution of the linear perturbations in this theory.

\section{Perturbation theory \label{perttheory}}

We will study the scalar perturbations of the theory (\ref{elp})-(\ref{idj}) around
the background (\ref{metric}). So, the background spatial metric is taken to be flat across all scales
comparable to the wavelength of the perturbations. The spatial harmonic functions
satisfying the equation $(\nabla^{2}+k^{2})Y=0$ are a complete set of the simple plane waves
\begin{equation}
Y(\vec{k},\vec{x})\propto e^{i\vec{k}\cdot\vec{x}}\,,\label{ked}
\end{equation}
where $\nabla^{2}=\delta^{ij}\partial_{i}\partial_{j}$, $k^{2}=\delta_{ij}k^{i}k^{j}$ and
$\vec{k}\cdot\vec{x}=\delta_{ij}k^{i}x^{j}$. In order to expand perturbations, scalars are expanded
by $Y$, while vectors and tensors are expanded respectively by
\begin{eqnarray}
&&Y_{i}=-\frac{1}{k}Y_{,i}=-i\frac{k_{i}}{k}Y\label{jet}\\
&&Y_{ij}=\frac{1}{k^{2}}Y_{,ij}+\frac{1}{3}\delta_{ij}Y=
\Big(\frac{1}{3}\delta_{ij}-\frac{k_{i}k_{j}}{k^{2}}\Big)Y\,.\label{ssk}
\end{eqnarray}
For a scalar perturbation the perturbed metric tensor $g_{\mu\nu}$ for a given wave-number $k$ is
generally parametrized in terms of four independent functions of time $A,B,H_{L},H_{T}$ as
\cite{Kodama:1985bj}
\begin{equation}
ds^{2}=g_{00}d\tau^{2}+2g_{0i}d\tau dx^{i}+g_{ij}dx^{i}dx^{j}\,,
\label{jer}
\end{equation}
where
\begin{eqnarray}
&&g_{00}=-a^{2}(1+2AY)\label{jet}\\
&&g_{0i}=-a^{2}BY_{i}\label{ecs}\\
&&g_{ij}=a^{2}(\delta_{ij}+2H_{L}Y\delta_{ij}+2H_{T}Y_{ij})\label{kwr}\,.
\end{eqnarray}
The perturbed scalar field is written as $\phi=\varphi(\tau)+\chi(\tau)Y$, where $\varphi$ is the
background field and $\chi$ the time dependent part of the perturbation.
The formulas for the perturbations of various geometric quantities as well as of the scalar field
derivatives are given in Appendix \ref{perturbations}.

The perturbations in the stress-energy tensor are decomposed into four components: density
$\delta\rho=\rho\delta=\rho\delta_{_{\!\!\!\!\sim}}Y$
%$\delta\!\!\!\slash(t) $
(with $\rho(\tau)$ the background density and $\delta_{_{\!\!\!\!\sim}}(\tau)$ the amplitude
of density perturbation), velocity $v$ (where the perturbed spatial velocity is $\frac{u^{i}}{u^{0}}=
vY^{i}$), isotropic pressure $\delta p=\varpi Y$ with $\varpi(\tau)$ measuring the amplitude of
isotropic pressure perturbation, and anisotropic stress $\frac{3}{2}(\rho+p)\sigma(\tau)$
in agreement with \cite{Ma:1994dv}. The perturbed stress-energy tensor takes the form
\begin{eqnarray}
 T^{0}_{\,\,\,0} &=& -\rho\PC{1 + \delta_{_{\!\!\!\!\sim}} Y} \nonumber\,,\\
 T^{0}_{\,\,\,i} &=& \PC{\rho + p}\PC{v - B}Y_{i} \nonumber\,,\\
 T^{i}_{\,\,\,j} &=& (p + \varpi Y)\delta^{i}_{j}
 + \frac{3}{2}\PC{\rho + p} \sigma Y^{i}_{\,\,\,j}\,.
\end{eqnarray}

The linearized version of the non-conservation law (\ref{idj}) is expressed as dynamical equations
for the energy density contrast $\delta_{_{\!\!\!\!\sim}}$ and the velocity $v$ as follows
\begin{widetext}
\begin{eqnarray}
&&\!\!\!\!\!\!\!\!\!\!\!\!\dot{\delta}_{_{\!\!\!\!\sim}} + (1+w)\big(kv+3\dot{H}_L\big)
+ 3 \mathcal{H}\Big(\frac{\delta p}{\delta\rho} - w\Big)\delta_{_{\!\!\!\!\sim}} =
\frac{\nu}{\varphi(\nu\!+\!8\pi\varphi^{2})}\dot{\chi}
-\frac{\nu(\nu\!+\!24\pi\varphi^{2})}{\varphi^{2}(\nu\!+\!8\pi\varphi^{2})^{2}}\dot{\varphi}\chi
\label{pertmatter1}\\
&&\!\!\!\!\!\!\!\!\!\!\!\!
\dot{v} - \dot{B} + (1\!-\!3w)\mathcal{H} (v\!-\!B) + \frac{\dot{w}}{1\!+\!w} (v\!-\!B)
- \frac{\delta p/ \delta \rho}{1\!+\!w} k \delta_{_{\!\!\!\!\sim}} - k A + k \sigma =
-\frac{\nu}{\varphi(\nu\!+\!8\pi\varphi^{2})}\frac{w}{1\!+\!w}k\chi\,, \label{pertmatter2}
\end{eqnarray}
\end{widetext}
where $w=p/\rho$ is the barotropic index of the background single fluid and
$\frac{\delta p}{\delta \rho}\,\delta_{_{\!\!\!\!\sim}}=\frac{\varpi}{\rho}$.
Note that the non-conservation Eq.~(\ref{idj}) for the background has been used in the
derivation of (\ref{pertmatter1}), (\ref{pertmatter2}).
We will consider that the evolution starts deep within matter domination and neglect radiation.
Hence, $\dot{w}$ will be set to zero and we will have just the matter component.

Next we proceed with the four perturbed field equations (\ref{elp}) which give
\newline
%\begin{flushleft}
  \underline{0 - 0 component}
%\end{flushleft}
\begin{eqnarray}
&&\varphi\PR{3\mathcal{H}^{2}A-k\mathcal{H}B-3\mathcal{H}\dot{H}_{L}-k^{2}\Big(H_L+\frac{H_T}{3}\Big)}
-\frac{3}{2}\mathcal{H}^{2}\chi=4\pi(-\rho a^{2} \delta_{_{\!\!\!\!\sim}}+\tau_{1})\label{jet}\\
&&\tau_{1}=\frac{24\pi\varphi^{2}\!-\!\nu}{2\lambda(\nu\!+\!8\pi\varphi^{2})^{3}}
\big[\nu\!+\!4\pi(2\!-\!3\lambda)\varphi^{2}\big]\dot{\varphi}^{2}\chi
+\frac{\varphi\dot{\varphi}}{\lambda(\nu\!+\!8\pi\varphi^{2})^{2}}
\Big{\{}\big[\nu\!+\!4\pi(2\!-\!3\lambda)\varphi^{2}\big](\dot{\varphi}A\!-\!\dot{\chi})
-4\pi(2\!-\!3\lambda)\varphi\dot{\varphi}\chi\Big{\}}\nn\\
&&\quad\quad
+\frac{6\nu\varphi\dot{\varphi}}{(\nu\!+\!8\pi\varphi^{2})^{2}}\mathcal{H}\chi
+\frac{\varphi^{2}}{\nu\!+\!8\pi\varphi^{2}}\PR{k^{2}\chi+3\mathcal{H}\dot{\chi}
-\dot{\varphi}\big(6\mathcal{H}A-kB-3\dot{H}_{L}\big)}\label{erj}
\end{eqnarray}
%\begin{flushleft}
  \underline{0 - $i$ component}
%\end{flushleft}
\begin{eqnarray}
&&\varphi\Big(\mathcal{H}A-\dot{H}_{L}-\frac{1}{3}\dot{H}_{T}\Big)=
\frac{4\pi\varphi^{2}}{\nu\!+\!8\pi\varphi^{2}}\big(\dot{\chi}-\mathcal{H}\chi-\dot{\varphi}A\big)
+\frac{4\pi\varphi}{\lambda(\nu\!+\!8\pi\varphi^{2})^{2}}
\big[\nu(1\!+\!\lambda)+4\pi(2\!-\!3\lambda)\varphi^{2}\big]\dot{\varphi}\chi\nn\\
&&\quad\quad\quad\quad\quad\quad\quad\quad\quad\quad\,\,\,\,
+\frac{4\pi}{k}(1\!+\!w)\rho a^{2}\big(v-B\big)
\label{wjr}
\end{eqnarray}
%\begin{flushleft}
  \underline{$i - j$ ($i \neq j$) component}
%\end{flushleft}
\begin{eqnarray}
&&\varphi\PR{-k^{2}A-k\big(\dot{B}+\mathcal{H}B\big)+\ddot{H}_{T}-k^{2}
\Big(H_{L}+\frac{H_{T}}{3}\Big)+\mathcal{H}\big(2\dot{H}_{T}-kB\big)}\nn\\
&&\quad\quad\quad\quad\quad\quad\quad\quad\quad
=\frac{8\pi\varphi^{2}}{\nu\!+\!8\pi\varphi^{2}}\big[k^{2}\chi
+\dot{\varphi}\big(kB-\dot{H}_{T}\big)\big]+12\pi(1\!+\!w)\rho a^{2}\sigma\label{wee}
\end{eqnarray}
%\begin{flushleft}
 \underline{$i - i$ component}
%\end{flushleft}
\begin{eqnarray}
&&\!\!\!\!\!
2\varphi\!\PR{\!\Big(\mathcal{H}^{2}\!+\!2\dot{\mathcal{H}}-\frac{k^{2}}{3}\Big)A
-\frac{k}{3}\big(\dot{B}\!+\!2\mathcal{H}B\big)+\mathcal{H}\dot{A}-\ddot{H}_{L}
\!-\!2\mathcal{H}\dot{H}_{L}-\frac{k^{2}}{3}\Big(H_{L}\!+\!\frac{H_{T}}{3}\Big)\!}-
\big(\mathcal{H}^{2}\!+\!2\dot{\mathcal{H}}\big)\chi=8\pi\big(a^{2}\varpi+\tau_{2}\big),\label{wet}\\
&&\!\!\!\!\!\!
\tau_{2}=\frac{\nu\!-\!24\pi\varphi^{2}}{2\lambda(\nu\!+\!8\pi\varphi^{2})^{3}}
\PR{(1\!+\!2\lambda)\nu\!+\!4\pi(2\!-\!3\lambda)\varphi^{2}}\!\dot{\varphi}^{2}\chi
+\frac{\varphi^{2}}{\nu\!+\!8\pi\varphi^{2}}\!\PR{\frac{2k}{3}\dot{\varphi}B\!-\!2(\ddot{\varphi}
\!+\!\mathcal{H}\dot{\varphi})A\!-\!\dot{\varphi}\dot{A}+\ddot{\chi}+\mathcal{H}\dot{\chi}+
\frac{2k^{2}}{3}\chi+2\dot{\varphi}\dot{H}_{L}}\nn\\
&&\,\,-\frac{\varphi\dot{\varphi}}{\lambda(\nu\!+\!8\pi\varphi^{2})^{2}}
\Big{\{}\big[(1\!+\!2\lambda)\nu\!+\!4\pi(2\!-\!3\lambda)\varphi^{2}\big](\dot{\varphi}A\!-\!\dot{\chi})
\!-\!4\pi(2\!-\!3\lambda)\varphi\dot{\varphi}\chi\Big{\}}
-\frac{2\nu\varphi}{(\nu\!+\!8\pi\varphi^{2})^{2}}\!\PR{\mathcal{H}\dot{\varphi}
\!+\!4\pi\lambda(3w\!-\!1)\rho a^{2}}\!\chi\label{iwt}.
\end{eqnarray}

Finally, the perturbed scalar field Eq.~(\ref{lrs}) gives
\newline
%\begin{flushleft}
\underline{$\delta \phi$ equation}
%\end{flushleft}
\begin{eqnarray}
\ddot{\chi}+2\mathcal{H}\dot{\chi}+k^{2}\chi-2\ddot{\varphi}A-\dot{\varphi}\big(4\mathcal{H}A+\dot{A}
-kB-3\dot{H}_{L}\big)=4\pi\lambda\Big(1\!-\!3\frac{\delta p}{\delta\rho}\Big)\rho a^{2}
\delta_{_{\!\!\!\!\sim}}\,.
\label{wbm}
\end{eqnarray}

Because of the Bianchi identities, not all the above equations are independent, but as for the
background, also here, one of these equations plays the role of the constraint. Therefore, one
equation is redundant and can be neglected. Since equation (\ref{wet}) is the most
complicated one containing also second derivatives, we will not make use of this in the numerical
analysis of Sec. \ref{lensingpot}. However, in Sec. \ref{subhorsec} of sub-horizon approximation,
Eq.~(\ref{wet}) will be used, while Eq.~(\ref{wjr}) will be the redundant one.

\subsection{Conformal Newtonian Gauge}

In the Newtonian gauge, one sets $H_{T}=B=0$, $A=\Psi$, $H_{L}=-\Phi$ \cite{Ma:1994dv}.
The matter equations of motion (\ref{pertmatter1}), (\ref{pertmatter2}) take the following form in
this gauge
\begin{widetext}
\begin{eqnarray}
&&\!\!\!\!\!\!\!\!\!\!\!\!\dot{\delta}_{_{\!\!\!\!\sim}} + (1+w)\big(kv-3\dot{\Phi}\big)
+ 3 \mathcal{H}\Big(\frac{\delta p}{\delta\rho} - w\Big)\delta_{_{\!\!\!\!\sim}} =
\frac{\nu}{\varphi(\nu\!+\!8\pi\varphi^{2})}\dot{\chi}
-\frac{\nu(\nu\!+\!24\pi\varphi^{2})}{\varphi^{2}(\nu\!+\!8\pi\varphi^{2})^{2}}\dot{\varphi}\chi
\label{pertmatter1conf}\\
&&\!\!\!\!\!\!\!\!\!\!\!\!
\dot{v} + (1\!-\!3w)\mathcal{H} v + \frac{\dot{w}}{1\!+\!w} v
- \frac{\delta p/ \delta \rho}{1\!+\!w} k \delta_{_{\!\!\!\!\sim}} - k \Psi + k \sigma =
-\frac{\nu}{\varphi(\nu\!+\!8\pi\varphi^{2})}\frac{w}{1\!+\!w}k\chi\,. \label{pertmatter2conf}
\end{eqnarray}
\end{widetext}
\noindent
The gravitational equations take the form
\newline
%\begin{flushleft}
  \underline{0 - 0 component}
%\end{flushleft}
\begin{eqnarray}
&&\varphi\big(3\mathcal{H}^{2}\Psi+3\mathcal{H}\dot{\Phi}+k^{2}\Phi\big)
-\frac{3}{2}\mathcal{H}^{2}\chi=4\pi(-\rho a^{2} \delta_{_{\!\!\!\!\sim}}+\tau_{1})\label{jeu}\\
&&\tau_{1}=\frac{24\pi\varphi^{2}\!-\!\nu}{2\lambda(\nu\!+\!8\pi\varphi^{2})^{3}}
\big[\nu\!+\!4\pi(2\!-\!3\lambda)\varphi^{2}\big]\dot{\varphi}^{2}\chi
+\frac{\varphi\dot{\varphi}}{\lambda(\nu\!+\!8\pi\varphi^{2})^{2}}
\Big{\{}\big[\nu\!+\!4\pi(2\!-\!3\lambda)\varphi^{2}\big](\dot{\varphi}\Psi\!-\!\dot{\chi})
-4\pi(2\!-\!3\lambda)\varphi\dot{\varphi}\chi\Big{\}}\nn\\
&&\quad\quad
+\frac{6\nu\varphi\dot{\varphi}}{(\nu\!+\!8\pi\varphi^{2})^{2}}\mathcal{H}\chi
+\frac{\varphi^{2}}{\nu\!+\!8\pi\varphi^{2}}\PR{k^{2}\chi+3\mathcal{H}\dot{\chi}
-\dot{\varphi}\big(6\mathcal{H}\Psi+3\dot{\Phi}\big)}\label{enj}
\end{eqnarray}
%\begin{flushleft}
  \underline{0 - $i$ component}
%\end{flushleft}
\begin{eqnarray}
\varphi\big(\mathcal{H}\Psi+\dot{\Phi}\big)=
\frac{4\pi\varphi^{2}}{\nu\!+\!8\pi\varphi^{2}}\big(\dot{\chi}-\mathcal{H}\chi-\dot{\varphi}\Psi\big)
+\frac{4\pi\varphi}{\lambda(\nu\!+\!8\pi\varphi^{2})^{2}}
\big[\nu(1\!+\!\lambda)+4\pi(2\!-\!3\lambda)\varphi^{2}\big]\dot{\varphi}\chi
+\frac{4\pi}{k}(1\!+\!w)\rho a^{2}v
\label{wor}
\end{eqnarray}
%\begin{flushleft}
  \underline{$i - j$ ($i \neq j$) component}
%\end{flushleft}
\begin{eqnarray}
&&\varphi\big(\Phi-\Psi\big)=\frac{8\pi\varphi^{2}}{\nu\!+\!8\pi\varphi^{2}}\chi
+\frac{12\pi}{k^2}(1\!+\!w)\rho a^{2}\sigma\label{wel}
\end{eqnarray}
%\begin{flushleft}
 \underline{$i - i$ component}
%\end{flushleft}
\begin{eqnarray}
&&\!\!\!\!\!\!\!\!\!\!\!\!\!\!\!\!\!\!\!\!\!
2\varphi\!\PR{\Big(\mathcal{H}^{2}\!+\!2\dot{\mathcal{H}}-\frac{k^{2}}{3}\Big)\Psi
+\frac{k^{2}}{3}\Phi+\ddot{\Phi}+2\mathcal{H}\dot{\Phi}+\mathcal{H}\dot{\Psi}}-
\big(\mathcal{H}^{2}\!+\!2\dot{\mathcal{H}}\big)\chi=8\pi\big(a^{2}\varpi+\tau_{2}\big),\label{wey}\\
&&\!\!\!\!\!\!\!\!\!\!\!\!\!\!\!\!\!\!\!\!\!
\tau_{2}=\frac{\nu\!-\!24\pi\varphi^{2}}{2\lambda(\nu\!+\!8\pi\varphi^{2})^{3}}
\PR{(1\!+\!2\lambda)\nu\!+\!4\pi(2\!-\!3\lambda)\varphi^{2}}\!\dot{\varphi}^{2}\chi
-\frac{\varphi^{2}}{\nu\!+\!8\pi\varphi^{2}}\!\PR{2(\ddot{\varphi}
\!+\!\mathcal{H}\dot{\varphi})\Psi+\dot{\varphi}\dot{\Psi}+2\dot{\varphi}\dot{\Phi}
-\ddot{\chi}-\mathcal{H}\dot{\chi}-\frac{2k^{2}}{3}\chi}\nn\\
&&\!\!\!\!\!\!\!\!\!\!\!\!\!-\frac{\varphi\dot{\varphi}}{\lambda(\nu\!+\!8\pi\varphi^{2})^{2}}
\Big{\{}\big[(1\!+\!2\lambda)\nu\!+\!4\pi(2\!-\!3\lambda)\varphi^{2}\big](\dot{\varphi}\Psi
\!-\!\dot{\chi})
\!-\!4\pi(2\!-\!3\lambda)\varphi\dot{\varphi}\chi\Big{\}}
-\frac{2\nu\varphi}{(\nu\!+\!8\pi\varphi^{2})^{2}}\!\PR{\mathcal{H}\dot{\varphi}
\!+\!4\pi\lambda(3w\!-\!1)\rho a^{2}}\!\chi\label{iws}.
\end{eqnarray}
Finally, the scalar field Eq.~(\ref{wbm}) becomes
\newline
%\begin{flushleft}
\underline{$\delta \phi$ equation}
%\end{flushleft}
\begin{eqnarray}
\ddot{\chi}+2\mathcal{H}\dot{\chi}+k^{2}\chi-2\ddot{\varphi}\Psi-\dot{\varphi}\big(4\mathcal{H}\Psi
+\dot{\Psi}+3\dot{\Phi}\big)=4\pi\lambda\Big(1\!-\!3\frac{\delta p}{\delta\rho}\Big)\rho a^{2}
\delta_{_{\!\!\!\!\sim}}\,.
\label{wou}
\end{eqnarray}

\subsection{Synchronous Gauge}

In this gauge, ones sets $A=B=0$, and $H_{L}=h/6$, $H_{T}=-3(\eta+h/6)$ \cite{Ma:1994dv}.
The matter equations of motion (\ref{pertmatter1}), (\ref{pertmatter2}) take the following form in
this gauge
\begin{widetext}
\begin{eqnarray}
&&\!\!\!\!\!\!\!\!\!\!\!\!\dot{\delta}_{_{\!\!\!\!\sim}} + (1+w)\Big(kv+\frac{\dot{h}}{2}\Big)
+ 3 \mathcal{H}\Big(\frac{\delta p}{\delta\rho} - w\Big)\delta_{_{\!\!\!\!\sim}} =
\frac{\nu}{\varphi(\nu\!+\!8\pi\varphi^{2})}\dot{\chi}
-\frac{\nu(\nu\!+\!24\pi\varphi^{2})}{\varphi^{2}(\nu\!+\!8\pi\varphi^{2})^{2}}\dot{\varphi}\chi
\label{pertmatter1syn}\\
&&\!\!\!\!\!\!\!\!\!\!\!\!
\dot{v} + (1\!-\!3w)\mathcal{H} v + \frac{\dot{w}}{1\!+\!w} v
- \frac{\delta p/ \delta \rho}{1\!+\!w} k \delta_{_{\!\!\!\!\sim}}  + k \sigma =
-\frac{\nu}{\varphi(\nu\!+\!8\pi\varphi^{2})}\frac{w}{1\!+\!w}k\chi\,. \label{pertmatter2syn}
\end{eqnarray}
\end{widetext}
One may remove the remaining freedom and completely define the coordinates by setting that cold
dark matter particles are at rest, having zero peculiar velocity $v$. Indeed, for cold dark matter
there is no stress, $\sigma=0$, and the isotropic pressure perturbation $\delta p$ should vanish as
well since pressure gradients should only be relevant at very small scales (and even then, this is
neglected sometimes). Thus, the condition of vanishing peculiar velocity, $v=0$, is consistent
with equation (\ref{pertmatter2syn}). Of course, such a result is not possible in Newtonian gauge
due to the presence of the gravitational potential $\Psi$ in (\ref{pertmatter2conf}).
\newline
The gravitational equations take the form
\newline
%\begin{flushleft}
  \underline{0 - 0 component}
%\end{flushleft}
\begin{eqnarray}
&&\varphi\big(\mathcal{H}\dot{h}-2k^{2}\eta\big)+3\mathcal{H}^{2}\chi
=8\pi(\rho a^{2} \delta_{_{\!\!\!\!\sim}}-\tau_{1})\label{jetn}\\
&&\tau_{1}=\frac{24\pi\varphi^{2}\!-\!\nu}{2\lambda(\nu\!+\!8\pi\varphi^{2})^{3}}
\big[\nu\!+\!4\pi(2\!-\!3\lambda)\varphi^{2}\big]\dot{\varphi}^{2}\chi
-\frac{\varphi\dot{\varphi}}{\lambda(\nu\!+\!8\pi\varphi^{2})^{2}}
\Big{\{}\big[\nu\!+\!4\pi(2\!-\!3\lambda)\varphi^{2}\big]\dot{\chi}
+4\pi(2\!-\!3\lambda)\varphi\dot{\varphi}\chi\Big{\}}\nn\\
&&\quad\quad
+\frac{6\nu\varphi\dot{\varphi}}{(\nu\!+\!8\pi\varphi^{2})^{2}}\mathcal{H}\chi
+\frac{\varphi^{2}}{\nu\!+\!8\pi\varphi^{2}}\Big(k^{2}\chi+3\mathcal{H}\dot{\chi}
+\frac{\dot{\varphi}}{2}\dot{h}\Big)\label{erjk}
\end{eqnarray}
%\begin{flushleft}
  \underline{0 - $i$ component}
%\end{flushleft}
\begin{eqnarray}
&&\varphi\dot{\eta}=
\frac{4\pi\varphi^{2}}{\nu\!+\!8\pi\varphi^{2}}\big(\dot{\chi}-\mathcal{H}\chi\big)
+\frac{4\pi\varphi}{\lambda(\nu\!+\!8\pi\varphi^{2})^{2}}
\big[\nu(1\!+\!\lambda)+4\pi(2\!-\!3\lambda)\varphi^{2}\big]\dot{\varphi}\chi
+\frac{4\pi}{k}(1\!+\!w)\rho a^{2}v
\label{wjrp}
\end{eqnarray}
%\begin{flushleft}
  \underline{$i - j$ ($i \neq j$) component}
%\end{flushleft}
\begin{eqnarray}
&&\varphi\PR{k^{2}\eta-6\mathcal{H}\Big(\dot{\eta}+\frac{\dot{h}}{6}\Big)
-3\ddot{\eta}-\frac{\ddot{h}}{2}}
=\frac{8\pi\varphi^{2}}{\nu\!+\!8\pi\varphi^{2}}\PR{k^{2}\chi
+3\dot{\varphi}\Big(\dot{\eta}+\frac{\dot{h}}{6}\Big)}+12\pi(1\!+\!w)\rho a^{2}\sigma\label{weey}
\end{eqnarray}
%\begin{flushleft}
 \underline{$i - i$ component}
%\end{flushleft}
\begin{eqnarray}
&&\!\!\!\!\!
\frac{\varphi}{3}\big(2k^{2}\eta-2\mathcal{H}\dot{h}-\ddot{h}\big)-
\big(\mathcal{H}^{2}\!+\!2\dot{\mathcal{H}}\big)\chi
=8\pi\big(a^{2}\varpi+\tau_{2}\big),\label{wep}\\
&&\!\!\!\!\!\!
\tau_{2}=\frac{\nu\!-\!24\pi\varphi^{2}}{2\lambda(\nu\!+\!8\pi\varphi^{2})^{3}}
\PR{(1\!+\!2\lambda)\nu\!+\!4\pi(2\!-\!3\lambda)\varphi^{2}}\!\dot{\varphi}^{2}\chi
+\frac{\varphi^{2}}{\nu\!+\!8\pi\varphi^{2}}\!\PC{\ddot{\chi}+\mathcal{H}\dot{\chi}+
\frac{2k^{2}}{3}\chi+\frac{\dot{\varphi}}{3}\dot{h}}\nn\\
&&\,\,\,\,\,\,\,+\frac{\varphi\dot{\varphi}}{\lambda(\nu\!+\!8\pi\varphi^{2})^{2}}
\Big{\{}\big[(1\!+\!2\lambda)\nu\!+\!4\pi(2\!-\!3\lambda)\varphi^{2}\big]\dot{\chi}
\!+\!4\pi(2\!-\!3\lambda)\varphi\dot{\varphi}\chi\Big{\}}
-\frac{2\nu\varphi}{(\nu\!+\!8\pi\varphi^{2})^{2}}\!\PR{\mathcal{H}\dot{\varphi}
\!+\!4\pi\lambda(3w\!-\!1)\rho a^{2}}\!\chi\label{iwm}.
\end{eqnarray}
Finally, the perturbed scalar field Eq.~(\ref{wbm}) gives
\newline
%\begin{flushleft}
\underline{$\delta \phi$ equation}
%\end{flushleft}
\begin{eqnarray}
\ddot{\chi}+2\mathcal{H}\dot{\chi}+k^{2}\chi+\frac{\dot{\varphi}}{2}\dot{h}
=4\pi\lambda\Big(1\!-\!3\frac{\delta p}{\delta\rho}\Big)\rho a^{2}\delta_{_{\!\!\!\!\sim}}\,.
\label{wom}
\end{eqnarray}

\section{The lensing potential \label{lensingpot}}

We assume the Newtonian gauge and neglecting anisotropic contributions from matter fields, $\sigma=0$,
the anisotropy Eq.~(\ref{wel}) yields the following algebraic relation between the gravitational
potentials and the scalar field perturbation
\begin{equation}
\Phi-\Psi=\frac{\chi}{D(\varphi)}\,,
\label{kqw}
\end{equation}
where we have set
\begin{equation}
D(\varphi)=\frac{\nu+8\pi\varphi^{2}}{8\pi\varphi}\,.
\label{kew}
\end{equation}
Equation (\ref{kqw}) defines the slip $\chi$ between the Newtonian potentials and expresses the departure
from standard general relativity where the anisotropy equation is the simple equation
$\Phi=\Psi$.

Since in (\ref{wor}) the only derivatives of the perturbed variables are encountered in the combination
$\dot{\Phi}-\frac{\dot{\chi}}{2D}$, defining $\Phi_{+}=\Phi-\frac{\chi}{2D}$, only the single
derivative $\dot{\Phi}_{+}$ will remain. Due to (\ref{kqw}) it is
\begin{equation}
\Phi_{+}=\frac{\Phi+\Psi}{2}\,,
\label{jet}
\end{equation}
which is called lensing potential. This is responsible for such effects as the integrated
Sachs-Wolfe effect in the CMB and weak lensing of distant galaxies.
Due to equation (\ref{kqw}), among the gravitational potentials $\Phi,\Psi$ and the scalar field
perturbation $\chi$, only two are independent quantities, which are given by $\Phi_{+},\chi$.
The variables $\Phi_{+},\chi$ are linear combinations of the gravitational potentials $\Phi,\Psi$,
and inversely
\begin{eqnarray}
&&\Phi=\Phi_{+}+\frac{\chi}{2D}\label{wrt}\\
&&\Psi=\Phi_{+}-\frac{\chi}{2D}\label{krh}\,.
\end{eqnarray}
We will transform the remaining gravitational equations (\ref{jeu}), (\ref{wor}) into a coupled
system of first-order differential equations for $\Phi_{+}$, $\chi$. These are the functions to be
evolved along with the perturbations of the matter fields. This analysis will facilitate the numerical
treatment of the equations and the interpretation of the results.

Starting with (\ref{wor}) we get, after the substitution (\ref{wrt}), (\ref{krh})
\begin{equation}
\Phi_{+}'=-\Big(1\!+\!\frac{\varphi'}{2D}\Big)\Phi_{+}+\frac{1}{2D^{2}}
\Big(D'\!+\!\frac{\varphi'}{2}\Big)\chi
+\frac{4\pi\varphi'}{\lambda(\nu\!+\!8\pi\varphi^{2})^{2}}
\big[\nu(1\!+\!\lambda)+4\pi(2\!-\!3\lambda)\varphi^{2}
\big]\chi+\frac{4\pi}{kH\varphi}(1\!+\!w)\rho a v\,,
\label{jir}
\end{equation}
where, as mentioned, a prime denotes differentiation with respect to $\ln{a}$.

A suitable linear combination of equations (\ref{jeu}) and (\ref{wor}) leads to an equation
containing the comoving density perturbation
\begin{equation}
\Delta=\delta_{_{\!\!\!\!\sim}}+\frac{3\mathcal{H}}{k}V\,,
\label{wit}
\end{equation}
where $V=(1+w)v$. Using again (\ref{wrt}), (\ref{krh}) to convert everything into
$\Phi_{+}$ and $\chi$, we finally get
\begin{eqnarray}
&&\!\!\!\!\!\!\!\!\!\!\frac{\varphi'}
{\lambda D}\chi'=-\frac{8\pi\rho}{H^{2}}\Delta
+3\chi
-\frac{4\pi(\nu\!-\!24\pi\varphi^{2})}{\lambda(\nu\!+\!8\pi\varphi^{2})^{3}}
\big[\nu\!+\!4\pi(2\!-\!3\lambda)\varphi^{2}\big]\varphi'^{2}\chi\nn\\
&&\!\!\!\!\!\!\!\!\!\!+\frac{8\pi}{\nu\!+\!8\pi\varphi^{2}}\Big{\{}\frac{3\nu\varphi'}{4\pi D}
+3\varphi^{2}\!+\!\frac{3\varphi^{2}\varphi'}{2D}
+\frac{2\varphi^{2}D'\varphi'}{2D^{2}}-\frac{2\!-\!3\lambda}{2\lambda D}\varphi\varphi'
\Big(\frac{\varphi\varphi'}{2D}\!+\!\varphi'\!+\!3\varphi\Big)\!-\!\frac{\nu\varphi'}{8\pi\lambda D}
\Big[\frac{\varphi'}{2D}+3(1\!+\!\lambda)\Big]\Big{\}}\chi\nn\\
&&\!\!\!\!\!\!\!\!\!\!
-\frac{3\varphi\varphi'}{D}\Phi_{+}'+\frac{8\pi\varphi'}{\nu\!+\!8\pi\varphi^{2}}
\Big(\frac{2\!-\!3\lambda}{2\lambda D}\varphi^{2}\varphi'\!+\!\frac{\nu}{8\pi\lambda D}
\varphi'\!-\!3\varphi^{2}\Big)\Phi_{+}
-\frac{2k^{2}\varphi}{a^{2}H^{2}}\Phi_{+}\,.
\label{jef}
\end{eqnarray}

We now have the tools to obtain the evolution of the linear perturbations in the complete Brans-Dicke
theory. Equations (\ref{pertmatter1conf}), (\ref{pertmatter2conf}) can easily be expressed in terms
of the lensing potential and primed derivatives. Equation (\ref{wey}) is the redundant equation.
We will solve numerically the system of the first-order differential equations
(\ref{pertmatter1conf}), (\ref{pertmatter2conf}), (\ref{jir}), (\ref{jef}). This system is
well-defined since the isotropic pressure $\delta p$ can be considered negligible in the scales
of interest. In principle $\delta p$ could be substituted from equation (\ref{wou}), however this
would bring unnecessary complexity due to the second derivatives of $\chi$, so we do not follow
this method. The remaining part of equation (\ref{wou}) can be checked for consistency in the end
for the numerical solution obtained. However, in the sub-horizon approximation of the next section
the consistency of this equation will become manifest.
In order to perform the numerical integration, we impose initial conditions on $\Phi_{+}$ and $\chi$
at a redsfhit of $z_{\rm{i}} = 1000$, as if we had minimal deviations from standard general relativity,
i.e. $\Phi_{+{\rm{i}}} = -1$ and $\chi_{\rm{i}}=0$. Since in GR the two Newtonian potentials remain
constant in the initial era of evolution and the lensing potential is a combination of these potentials,
it is reasonable to set $\Phi_{+\rm{i}}'=0$. Then, Eqs.~(\ref{jir}), (\ref{jef}) provide the initial
velocity $v_{\rm{i}}$ of the matter perturbation and the initial comoving density perturbation
$\Delta_{\rm{i}}$. Indeed, using (\ref{eji}) initially, we get the standard GR relations
\begin{eqnarray}
v_{\rm{i}}&=&\frac{2 k}{3a_{\rm{i}}H_{\rm{i}}}\Phi_{+\rm{i}}
\label{init_pert_conds1}\\
\Delta_{\rm{i}}&=& -\frac{2 k^2}{3a_{\rm{i}}^2H_{\rm{i}}^2}\Phi_{+\rm{i}}\,.
\label{init_pert_conds}
\end{eqnarray}

\begin{figure}[t!]
\begin{minipage}{.5\linewidth}
\centering
\subfloat[]{\label{main:a}\includegraphics[scale=.41]{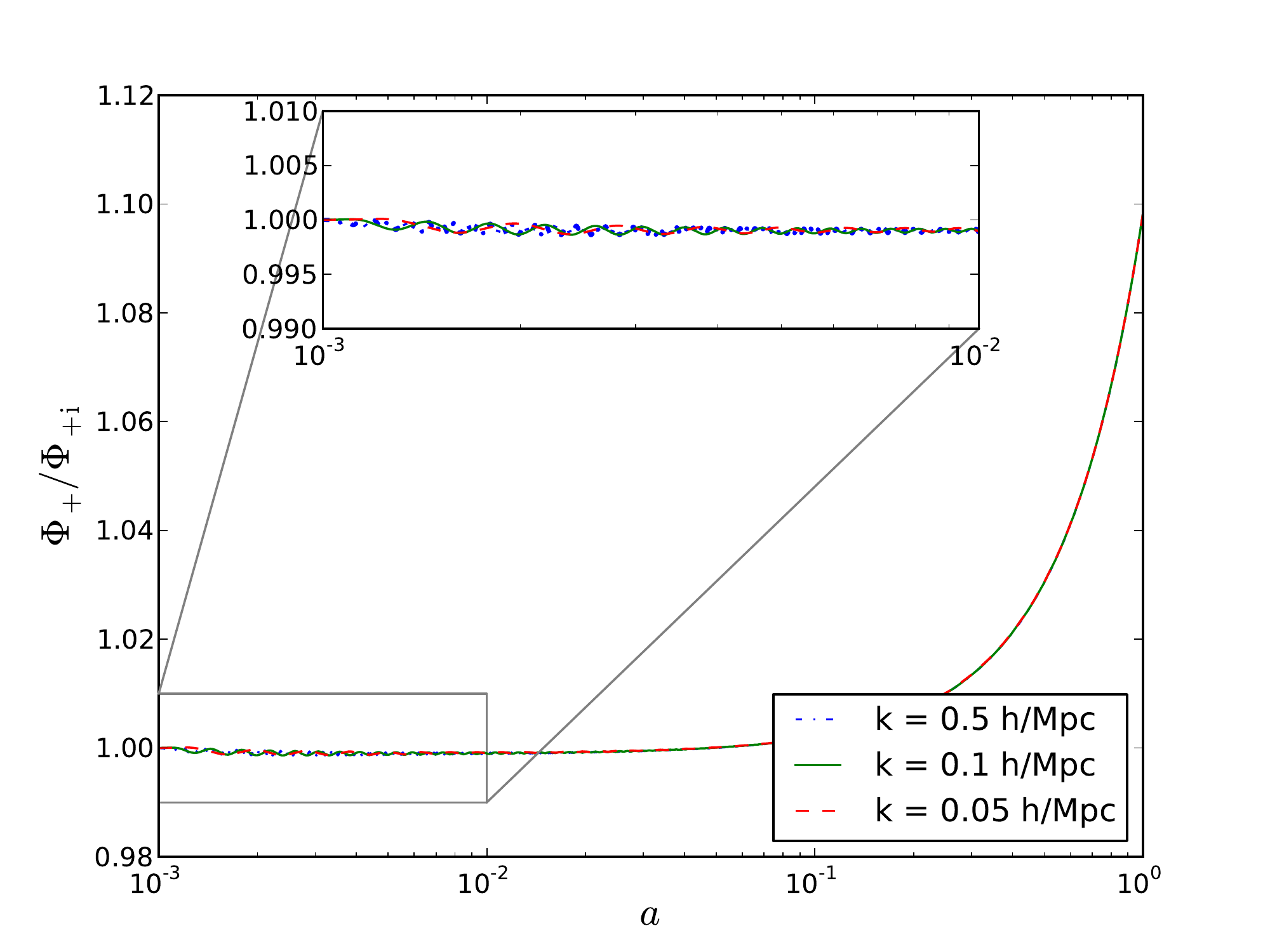}}
\end{minipage}%
\begin{minipage}{.5\linewidth}
\centering
\subfloat[]{\label{main:b}\includegraphics[scale=.41]{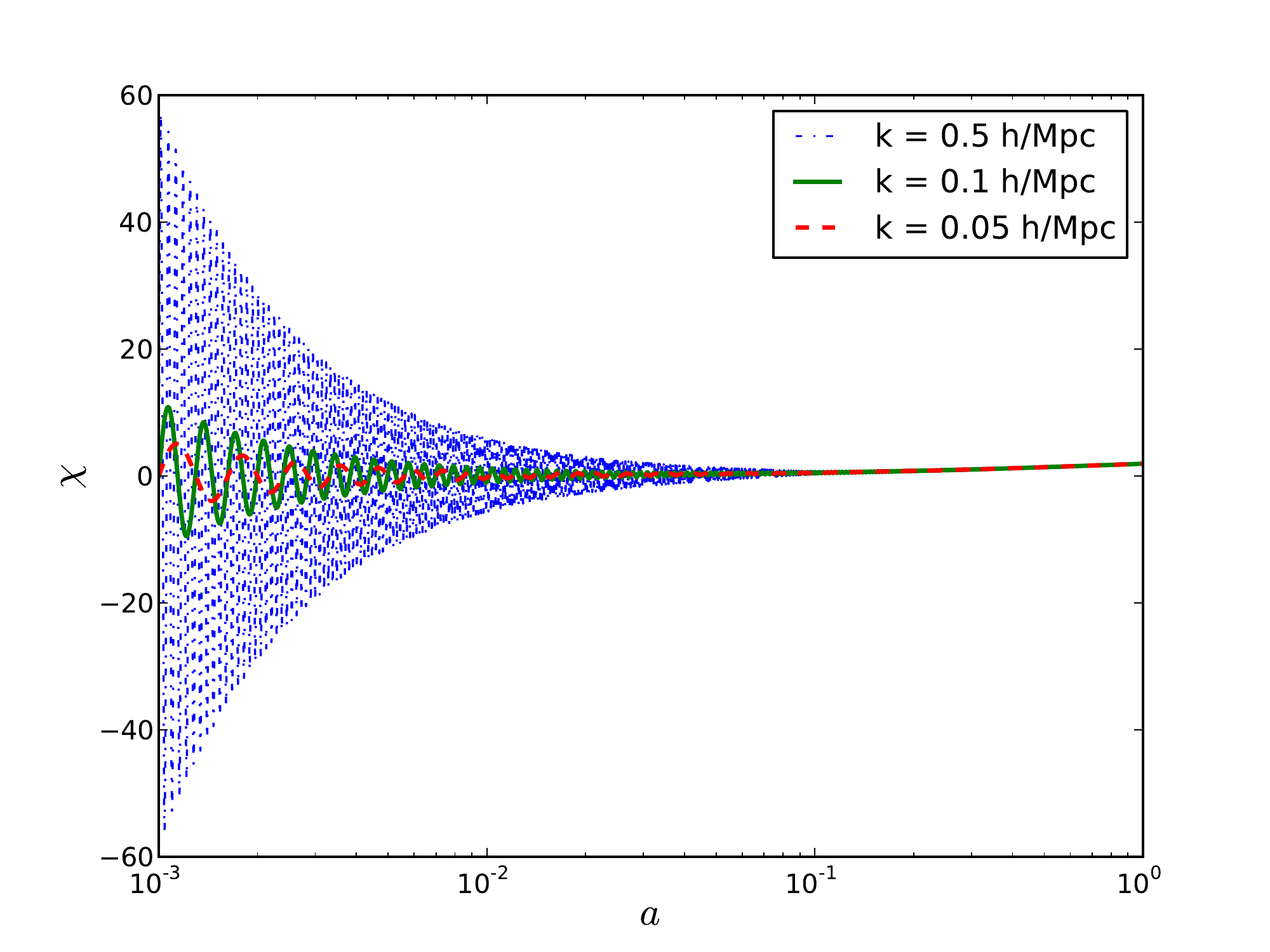}}
\end{minipage}
\caption{\label{perturbations_figure}We plot the evolution of the lensing potential $\Phi_{+}$ and
the slip $\chi$ between the Newtonian potentials as a function of scale factor for different $k$ (h/Mpc)
scales, using the CBD model.}
\end{figure}

We can see the evolution of the lensing potential $\Phi_{+}$ and the slip $\chi$ in
Figs.~\ref{perturbations_figure} (a) and (b), respectively, for the CBD model.
The first immediate observation is that the evolution of the linear
perturbations in the complete Brans-Dicke theory (both the lensing potential and the slip)
is scale-dependent, particularly at early-times. This is a general feature of modified gravity
theories, and is in complete contrast to the scale-independent GR+$\Lambda$CDM predictions.
Indeed in GR, the absence of $\chi$ makes equations (\ref{jir}) and (\ref{pertmatter2conf}) an autonomous
system for $\Phi,\hat{v}=v/k$, without containing $k$. Together with the above initial conditions
$\Phi_{\rm{i}}=-1$, $\hat{v}_{\rm{i}}=2\Phi_{\rm{i}}/(3\mathcal{H}_{\rm{i}})$, which do not contain $k$
as well, it arises that $\Phi,\hat{v}$ are scale-independent in GR, while (\ref{jef}) shows that
$\delta_{_{\!\!\!\!\sim}}$ is scale-dependent in GR. In CBD however, the existence of the last
$k$-term in (\ref{jef}), as well as the $kv$ term in (\ref{pertmatter1conf}), result to the
scale-dependence of all perturbations.

Then, we have the oscillatory behavior of the slip between the Newtonian potentials,
shown in Fig.~\ref{perturbations_figure} (b). This is also
observed in other (non-interacting) scalar-tensor theories such as metric $f(R)$ \cite{frperturbed},
or the hybrid metric-Palatini theory \cite{hybridpert}, and can be understood from Eq.~(\ref{wou})
which is the equation of a damped harmonic oscillatory with a driving term. We observe these
oscillations mainly at early-times, and they become more pronounced the smaller the scales
(higher $k$'s) we consider, as these modes start deep within the range of action of the additional
force the scalar degree of freedom mediates. As in standard Brans-Dicke theory (with at most a
constant potential), we have a massless scalar field. Hence, its effective Compton radius can include
and impact even the largest scales (smallest $k$) we consider at early times.
These oscillations could lead to instabilities at early-times. For instance, in
metric $f(R)$, the oscillations in the gravitational potentials manifest in the perturbation of
the metric Ricci scalar $\delta R$, leading to a possible overproduction of new massive scalar
particles in the very early Universe \cite{starobinskyfr}. A more detailed study on this is,
however, beyond the scope of the current work.

As the cosmological evolution continues, the oscillations in $\chi$ get progressively damped by
the Hubble friction term in Eq.~(\ref{wou}), and they eventually get smoothed and unnoticeable
toward the present. As we approach $a = 1$, we see that the equilibrium position of $\chi$ is
shifted from zero to a positive value, and shows a tendency to increase. This is due to the driving
term in Eq.~(\ref{wou}), which tries to displace $\chi$ from the equilibrium position set by the
initial conditions. Hence, as $\delta_{_{\!\!\!\!\sim}}$ grows, the driving term will become more
important, and its overall effect will be more significant the larger the value of $\lambda$ is.

Lastly, we have the lensing potential $\Phi_{+}$ in Fig.~\ref{perturbations_figure} (a). We have
again the distinct effect of the scale-dependent oscillations at early times that propagate from
the evolution in $\chi$. Such rapid oscillations in $\Phi_{+}$ can contribute to a significant
early-times integrated Sachs-Wolfe (eISW) effect, which could impact the CMB as seen in
Ref.~\cite{earlymgcons} for instance.  Furthermore, we also have a noticeable departure from
standard GR+$\Lambda$CDM toward the present. We can see in Fig.~\ref{perturbations_figure} (a)
that the absolute value of the lensing potential exhibits a distinct growing
tendency at late-times. In the standard cosmological model, with the onset of cosmic acceleration,
the lensing potential decays due to the expanding background. What we see in CBD is that, despite
having accelerating background solutions, $|\Phi_{+}|$ actually grows as we approach $a = 1$, yielding
a late-times integrated Sachs-Wolfe (lISW) effect opposite to that of $\Lambda$CDM. This should
produce a noticeable impact on the larger scales of the CMB, which might cause difficulties with
current observations of the lISW \cite{lisw1,lisw2,lisw3}.

Note also from Eqs.~(\ref{wrt}), (\ref{krh}) and Fig.~\ref{perturbations_figure} that the Newtonian
potentials $\Phi,\Psi$ oscillate around $-1$ at early times. Although these potentials normally
acquire negative values due to the attractive character of gravity, however, it can be seen that
$\Psi$ passes to positive values at late times. This late-times behaviour of $\Psi$ will become
significant in Sec.~\ref{sec_growth}, where the behaviour of $\delta_{_{\!\!\!\!\sim}}$ will be studied.

\section{Sub-horizon approximation \label{subhorsec}}

We now consider wavemodes that are deep within the Hubble radius such that $k \gg aH$. In this limit,
we adopt the quasistatic approximation, discarding time derivatives of perturbations when compared to
their spatial variation. This is generally a good approximation for scalar-tensor theories on small
scales~\cite{lombriser:15a}. In practice, this allows one to keep the terms proportional to
$k^2/(a^{2}H^{2})$, as well as those related to the matter perturbation $\delta_{_{\!\!\!\!\sim}}$ and
$\varpi$, and is known as the sub-horizon approximation \cite{sub1,sub2}. Equation (\ref{jeu}) becomes
\begin{equation}
\Phi=-\frac{4\pi}{\varphi}\,\frac{a^{2}}{k^{2}}\rho\delta_{_{\!\!\!\!\sim}}+
\frac{4\pi\varphi}{\nu\!+\!8\pi\varphi^{2}}\chi\,.
\label{jer}
\end{equation}
Therefore, Eq.~(\ref{jeu}), which gave the differential equation (\ref{jef}) for $\chi$, has now
become an algebraic equation. Equation (\ref{wel}) coincides with
equation (\ref{kqw}) for the slip. Also the complicated equation (\ref{wey}) gets the simple
algebraic form
\begin{equation}
\Phi-\Psi
=\frac{8\pi\varphi}{\nu\!+\!8\pi\varphi^{2}}\chi+\frac{12\pi}{\varphi}\,\frac{a^{2}}{k^{2}}\varpi\,.
\label{grw}
\end{equation}
Due to Eq.~(\ref{kqw}), Eq.~(\ref{grw}) gives $\varpi=0$, so in the sub-horizon approximation
our previous assumption of negligible isotropic pressure perturbation is verified.
This means that in the context of the present approximation the $i-i$ and $i-j$ ($i\neq j$)
equations coincide if $\varpi=0$, and they provide information for the potential $\Psi$.
The differential equation (\ref{wou}) becomes
\begin{equation}
\chi=4\pi\lambda\frac{a^{2}}{k^{2}}\rho\delta_{_{\!\!\!\!\sim}}-12\pi\lambda\frac{a^{2}}{k^{2}}\varpi\,,
\label{sek}
\end{equation}
where $\varpi=0$ has to be set. Therefore, we see a proportionality between the slip $\chi$ and
the matter perturbation $\delta_{_{\!\!\!\!\sim}}$. Finally, equation (\ref{wor}) or the
lensing potential equation (\ref{jir}) does not accept any simplification and is the redundant equation
in this approximation, which should be satisfied on-shell.

From Eqs.~(\ref{jer}), (\ref{sek}) we can express $\Phi$ in terms of $\delta_{_{\!\!\!\!\sim}}$ as
\begin{equation}
\frac{k^2}{a^2}\Phi=-\frac{4\pi}{\varphi}\PC{\frac{\nu+8\pi\varphi^2\PC{1-\lambda/2}}
{\nu+8\pi \varphi^2}}\rho\delta_{_{\!\!\!\!\sim}}\,.
\label{lqr}
\end{equation}
Then, from equation (\ref{grw}) we get $\Psi$ as
\begin{equation}
\frac{k^2}{a^2}\Psi=-\frac{4\pi}{\varphi}\PC{\frac{\nu+8\pi\varphi^2\PC{1+\lambda/2}}
{\nu+8\pi\varphi^2}}\rho\delta_{_{\!\!\!\!\sim}}\,,
\label{pwe}
\end{equation}
and we write again Eq.~(\ref{sek}) for completeness
\begin{equation}
\chi=4\pi\lambda\frac{a^{2}}{k^{2}}\rho\delta_{_{\!\!\!\!\sim}}\,.
\label{sekc}
\end{equation}
Equations (\ref{lqr}), (\ref{pwe}), (\ref{sekc}) express algebraically the gravitational and scalar
field perturbations in terms of the matter density perturbation $\delta_{_{\!\!\!\!\sim}}$. This
$\delta_{_{\!\!\!\!\sim}}$ is given by the system of equations (\ref{pertmatter1conf}),
(\ref{pertmatter2conf}) after substitution of (\ref{lqr}), (\ref{pwe}), (\ref{sekc}).
In the next section, we will derive an autonomous second-order differential equation for
$\delta_{_{\!\!\!\!\sim}}$ within the sub-horizon approximation. Note also from Eqs.~(\ref{lqr}),
(\ref{pwe}), (\ref{sekc}) the proportionality of $\Phi,\Psi$ to $\chi$, in agreement with the
late-times behaviours derived numerically in the previous section without any approximation.

Lastly, we can relate from the above expressions the lensing potential $\Phi_{+}$
in the sub-horizon approximation to the scalar-field perturbation as
\begin{equation}
\Phi_{+}=-\frac{1}{\lambda\varphi}\chi\,.
\label{jeufv}
\end{equation}
This equation also coincides with the sub-horizon limit of the slip equation (\ref{jef}), where
a term proportional to $v/k$ should be ignored in this limit (this is due to that ignoring
the various terms in (\ref{jef}) means from (\ref{jir}) ignoring the term $v/k$).
From Eq.~(\ref{jeufv}) we can anticipate that if $\chi$ grows at late-times, then $\Phi_{+}$ will
follow that behavior, increasing in absolute amplitude, in agreement with Fig.~\ref{perturbations_figure}
(a). The scale-independence of $\chi$ at late-times, shown in Fig.~\ref{perturbations_figure} (b)
and explained in the next section, implies also the same independence for $\Phi_{+},\Phi,\Psi$.
This is consistent with the non-massive Brans-Dicke theory
\cite{defelice}, and contrasts, for instance, with metric $f(R)$ theories \cite{frperturbed}.
This is the reason why we are not able to resolve the differences in the evolution of the
lensing potential and $\chi$ in Fig.~\ref{perturbations_figure} for the different $k$ scales when
we approach the present time.

We can now write the two functions that are commonly used to parametrize deviations from general
relativity in modified theories of gravity, $\mu(a,k)$ and $\gamma(a,k)$. The former defines the
relation between the Newtonian potential $\Psi$ and the matter density perturbation, while the
latter parametrizes the ratio between the gravitational potentials, such as \cite{subparam1,subparam2}
\begin{eqnarray}
\frac{k^2}{a^2} \Psi &=& -4 \pi \mu(a,k) \rho\delta_{_{\!\!\!\!\sim}}\,, \noindent\\
\frac{\Phi}{\Psi} &=& \gamma(a,k)\,.
\end{eqnarray}
Hence, for the complete Brans-Dicke theory, these functions will take the form
\begin{eqnarray}{\label{cbdparam}}
\mu_{\rm{CBD}}(a,k) &=& \frac{1}{\varphi}\frac{\nu + 8\pi\varphi^2\PC{1 + \lambda/2}}{\nu + 8\pi\varphi^2}\,, \\
\gamma_{\rm{CBD}}(a,k) &=& \frac{\nu + 8 \pi \varphi^2 \PC{1 - \lambda/2}}{\nu + 8 \pi \varphi^2 \PC{1 + \lambda/2}}\,,
\end{eqnarray}
which recover the known results for standard massless Brans-Dicke in the limit of $\nu = 0$
\cite{defelice}
\begin{eqnarray}
\mu_{\nu = 0}(a,k) &=&\frac{2 \omega_{\rm{BD}} + 4}{2 \omega_{\rm{BD}} + 3}\,\frac{1}{\varphi}\,, \\
\gamma_{\nu = 0}(a,k) &=& \frac{\omega_{\rm{BD}}+1}{\omega_{\rm{BD}}+2}\,.
\end{eqnarray}

\section{Growth rate\label{sec_growth}}

The equations for matter perturbations (\ref{pertmatter1conf}), (\ref{pertmatter2conf}) in the matter
era with $\sigma=\varpi=0$ take the following form (without making use of the approximation on
sub-horizon scales)
\begin{eqnarray}
&&\dot{\delta}_{_{\!\!\!\!\sim}} + kv-3\dot{\Phi}=
\frac{\nu}{\varphi(\nu\!+\!8\pi\varphi^{2})}\dot{\chi}
-\frac{\nu(\nu\!+\!24\pi\varphi^{2})}{\varphi^{2}(\nu\!+\!8\pi\varphi^{2})^{2}}\dot{\varphi}\chi
\label{ejr}\\
&&\dot{v} + \mathcal{H} v - k \Psi=0\,.
\label{kpw}
\end{eqnarray}
We differentiate (\ref{ejr}) once more and eliminate $\dot{v},v$ from (\ref{ejr}), (\ref{kpw})
to find the equation of motion for $\delta_{_{\!\!\!\!\sim}}$
\begin{equation}
\ddot{\delta_{_{\!\!\!\!\sim}}}+\mathcal{H}\dot{\delta_{_{\!\!\!\!\sim}}}+k^{2}\Psi
-3\ddot{\Phi}-3\mathcal{H}\dot{\Phi}=\frac{1}{a}
\Big[\frac{\nu a}{\varphi(\nu\!+\!8\pi\varphi^{2})}\dot{\chi}
-\frac{\nu(\nu\!+\!24\pi\varphi^{2})a}{\varphi^{2}(\nu\!+\!8\pi\varphi^{2})^{2}}\dot{\varphi}\chi\Big]
^{\LargerCdot}\label{lke}\,.
\end{equation}
Now the sub-horizon approximation can be implemented and equation (\ref{lke}) gets simplified as
\begin{equation}
\ddot{\delta_{_{\!\!\!\!\sim}}}+\mathcal{H}\dot{\delta_{_{\!\!\!\!\sim}}}+k^{2}\Psi=0\,.
\label{jwkj}
\end{equation}
Converting to e-folding time $\ln{a}$ we get
\begin{equation}
\delta_{_{\!\!\!\!\sim}}''+\Big(\frac{\mathcal{H}'}{\mathcal{H}}+1\Big)\delta_{_{\!\!\!\!\sim}}'
+\frac{k^{2}}{\mathcal{H}^{2}}\Psi=0\,.
\label{lkwr}
\end{equation}
Using equation (\ref{pwe}) to replace $\Psi$ we obtain
\begin{equation}
\delta_{_{\!\!\!\!\sim}}''+\Big(\frac{\mathcal{H}'}{\mathcal{H}}+1\Big)\delta_{_{\!\!\!\!\sim}}'
-\frac{4\pi}{\mathcal{H}^{2}}\,\frac{\nu\!+\!4\pi(2\!+\!\lambda)\varphi^{2}}
{\varphi(\nu\!+\!8\pi\varphi^{2})}\rho a^{2}\delta_{_{\!\!\!\!\sim}}=0\,.
\label{dhd}
\end{equation}
The second order differential equation (\ref{dhd}) for the dynamics of the linear matter
perturbations $\delta_{_{\!\!\!\!\sim}}$ can also be written as
\begin{equation}
f'+f^{2}+\Big(\frac{\mathcal{H}'}{\mathcal{H}}+1\Big)f
-\frac{4\pi}{\mathcal{H}^{2}}\,\frac{\nu\!+\!4\pi(2\!+\!\lambda)\varphi^{2}}
{\varphi(\nu\!+\!8\pi\varphi^{2})}\rho a^{2}=0\,,
\label{jlde}
\end{equation}
where $f=\frac{d\ln{\delta_{_{\!\!\!\!\sim}}}}{d\ln{a}}$ is the linear growth rate.
The modified gravitational coupling predicted by the CBD theory through
$\mu_{\rm{CBD}}(a,k)$ will lead to a growth history that is different than that of an effective
dark energy model within GR that exhibits the same expansion history as our complete Brans-Dicke model.

Equation (\ref{dhd}) defines an autonomous differential equation for the quantity
$\hat{\delta}_{_{\!\!\!\!\sim}}=\delta_{_{\!\!\!\!\sim}}/k^{2}$. No $k$-dependence is present
in this equation for $\hat{\delta}_{_{\!\!\!\!\sim}}$. Moreover, from the initial conditions
(\ref{init_pert_conds1}), (\ref{init_pert_conds}) we get $\hat{\delta}_{_{\!\!\!\!\sim}\rm{i}}=
-2\Phi_{+\rm{i}}/(3\mathcal{H}_{\rm{i}}^{2})$, which also does not depend on $k$. This initial condition
also arises from the sub-horizon Eqs.~(\ref{jeufv}), (\ref{sekc}) with the use of (\ref{eji}).
Since both the differential equation and the initial condition of $\hat{\delta}_{_{\!\!\!\!\sim}}$
do not depend on $k$, thus $\hat{\delta}_{_{\!\!\!\!\sim}}$ is scale-independent, which means that
$\delta_{_{\!\!\!\!\sim}}$ is proportional to $k^{2}$ in the sub-horizon limit (the same is true
in the sub-horizon limit of GR). From equation (\ref{sekc}), we obtain that $\chi$ is
scale-independent in this approximation, in agreement with the late-times behaviour of
Fig.~\ref{perturbations_figure} (b). Thus, all $\Phi_{+},\Phi,\Psi$ are scale-independent in this limit.
Finally, since knowing $\hat{\delta}_{_{\!\!\!\!\sim}}$ means from (\ref{pwe}) that $\Psi$ is known
and scale-independent, thus Eq.~(\ref{kpw}) is converted into an autonomous differential equation for
$\hat{v}=v/k$ which does not depend on $k$. Additionally, the initial condition (\ref{init_pert_conds1})
is $\hat{v}_{\rm{i}}=2\Phi_{+\rm{i}}/(3\mathcal{H}_{\rm{i}})$, which also does not depend on $k$.
Therefore, $\hat{v}$ is scale-independent and $v$ depends linearly on $k$ in the
sub-horizon approximation (this is also true in GR, but at all times).

\begin{figure}[t!]
\begin{center}
\includegraphics[scale = 0.50]{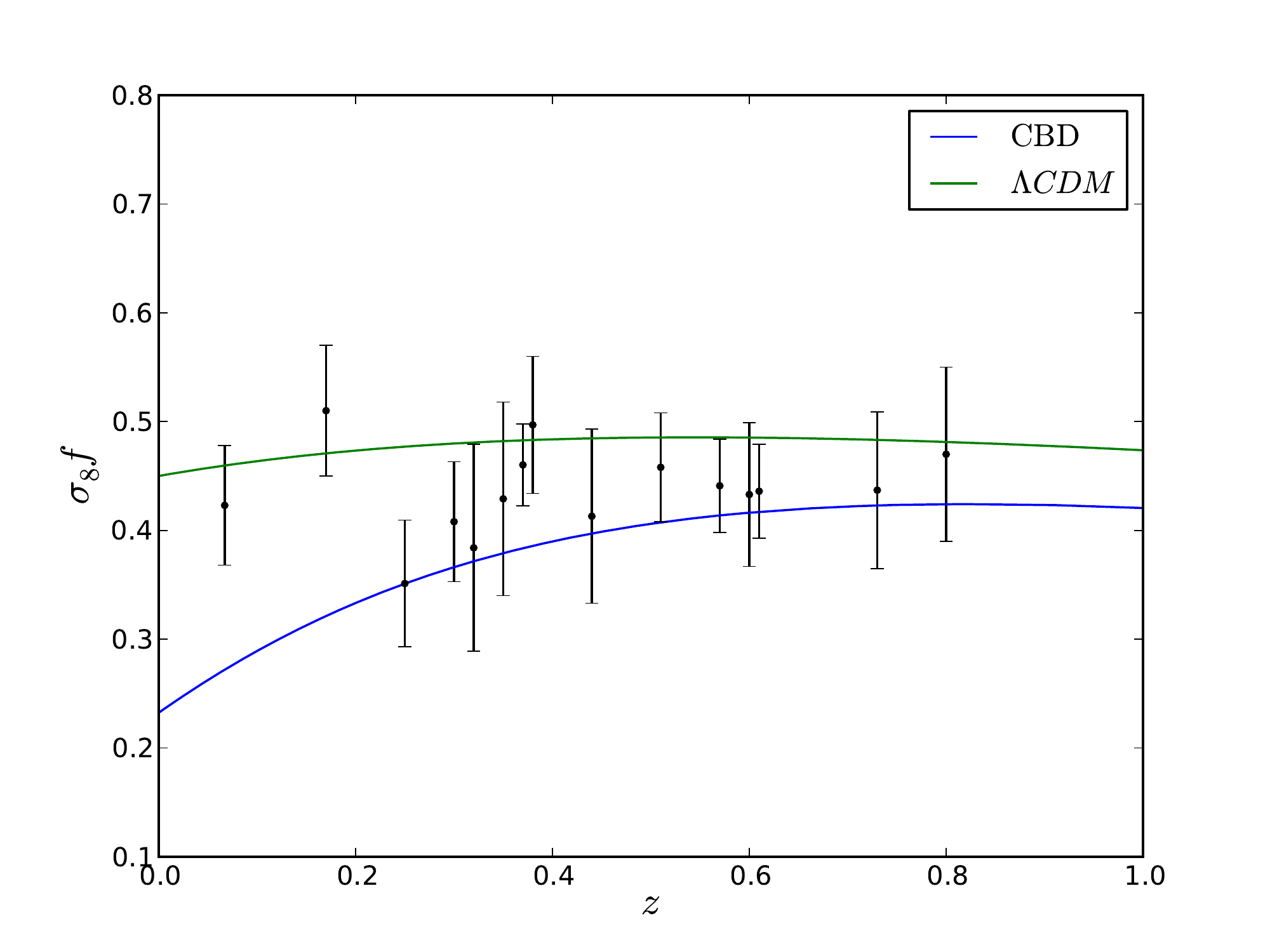}
\caption{\label{growth_rate} We plot $f\sigma_8$ for the CBD model against $\Lambda$CDM.
The data points used can be seen in Table~\ref{table_growth}. The parameters used were $\lambda=1$,
$\nu=-100$ and $\hat{\Omega}_{\rm{m}}=0.17$, $\varphi_{\rm{i}}=0.029$. We have taken
$\sigma_{8}^{0} = 0.83$, as measured by the Planck collaboration \cite{planckparams}.}
\end{center}
\end{figure}

In Fig.~\ref{growth_rate} we plot for recent redshifts the numerical evolution of $f\sigma_{8}(z)$,
also known as growth rate, with the amplitude of fluctuations $\sigma_{8}(z)$ given by
\begin{equation}{\label{sigma8}}
\sigma_{8}(z) = \sigma_{8}^{0} \frac{\delta_{_{\!\!\!\!\sim}}(z,k)}{\delta_{_{\!\!\!\!\sim}}(0,k)},
\end{equation}
where the current value of $\sigma_{8}$ can be estimated through the cosmic microwave
background \cite{planckparams}, weak-lensing \cite{sigma8_wl} or galaxy clustering \cite{sigma8_gal}.
On the other hand, $f\sigma_8$ can be extracted from redshift space distortions (RSD) observations as
a function of redshift. The most recent $f\sigma_8$ data points available were used in
Fig.~\ref{growth_rate}, and can be consulted in Table~\ref{table_growth}.
The plot of Fig.~\ref{growth_rate} has been made using the exact equations discussed in
Sec.~\ref{lensingpot}, and not the approximated equation (\ref{jlde}).

\begin{table}[t]
\centering{} %
\begin{tabular}{|c|c|c|c|}
\hline
Survey & $z$  & $\sigma_{8}f(z)$  & Source\tabularnewline
\hline
\hline
6dFGRS  & $0.067$  & $0.423\pm0.055$  & Beutler et al. (2012)~\cite{growth_1}\tabularnewline
\hline
\multirow{2}{*}{LRG-200} & $0.25$  & $0.3512\pm0.0583$  & \multirow{2}{*}{Samushia et al. (2012)~\cite{growth_2}}\tabularnewline
\cline{2-3}
 & $0.37$  & $0.4602\pm0.0378$  & \tabularnewline
\hline
\multirow{5}{*}{BOSS} & $0.30$  & $0.408\pm0.0552$  & \multirow{2}{*}{Tojeiro et al. (2012)~\cite{growth_3}}\tabularnewline
\cline{2-3}
 & $0.60$  & $0.433\pm0.0662$  & \tabularnewline
\cline{2-4}
 & $0.38$  & $0.497\pm0.063$  & \multirow{3}{*}{Alam et al. (2016) \cite{growth_4}}\tabularnewline
\cline{2-3}
 & $0.51$  & $0.458\pm0.050$  & \tabularnewline
\cline{2-3}
 & $0.61$  & $0.436\pm0.043$  & \tabularnewline
\hline
\multirow{2}{*}{WiggleZ } & $0.44$  & $0.413\pm0.080$  & \multirow{2}{*}{Blake (2011) \cite{growth_5}}\tabularnewline
\cline{2-3}
 & $0.73$  & $0.437\pm0.072$  & \tabularnewline
\hline
Vipers  & $0.8$  & $0.47\pm0.08$  & De la Torre et al. (2013) \cite{growth_6}\tabularnewline
\hline
2dFGRS  & $0.17$  & $0.51\pm0.06$  & Percival et al. (2004) \cite{growth_7,growth_8}\tabularnewline
\hline
LRG  & $0.35$  & $0.429\pm0.089$  & Chuang and Wang (2013) \cite{growth_9}\tabularnewline
\hline
LOWZ  & $0.32$  & $0.384\pm0.095$  & Chuang et al. (2013) \cite{growth_10}\tabularnewline
\hline
CMASS  & $0.57$  & $0.441\pm0.043$  & Samushia et al. (2013) \cite{growth_11}\tabularnewline
\hline
\end{tabular}\protect\protect\caption{\label{table_growth} RSD $f \sigma_8$ measurements from
various sources, used in Fig.~\ref{growth_rate}.}
\end{table}

We can see in this figure that the CBD theory predicts less growth than $\Lambda$CDM, and
could potentially provide a better fit to existent RSD data than the concordance cosmological model.
This may seem counter-intuitive given that, in Sec.~\ref{lensingpot}, we concluded that the
lensing potential $\Phi_{+}$ exhibited a distinct late-time growth as the slip between the
gravitational potentials also grew at late-times.
Since the scale used in Fig.~\ref{growth_rate} is certainly sub-horizon at low redshifts, the
decrease of $f\sigma_8$ or also of $f$, compared to $\Lambda$CDM, can be explained from the
last term in Eq.~(\ref{jlde}). The effective gravitational coupling concerning the perturbations
is given from (\ref{jlde}) as $G_{\rm{eff}}=\frac{\nu+4\pi(2+\lambda)\varphi^{2}}
{\varphi(\nu+8\pi\varphi^{2})}$, and it can be seen that $G_{\rm{eff}}$ passes from positive to
negative values recently. This change of sign happens when the scalar field crosses the critical
value $\varphi_{\rm{c}}=\sqrt{|\nu|/[4\pi(2\!+\!\lambda)]}$, and it is
$\varphi_{\rm{c}}<\varphi_{\infty}$ as long as $\lambda>0$.
For $\Lambda$CDM or for BD, the corresponding $G_{\rm{eff}}$'s are positive.
Therefore, as $\varphi$ grows toward the present, $f'$ in the CBD theory acquires a negative
contribution (or before that, a decaying positive contribution) due to $G_{\rm{eff}}$, providing
less growth. Similarly, the equation governing
$f\sigma_{8}$ in the sub-horizon approximation arises from (\ref{jlde}) as
\begin{equation}{\label{fsigma8}}
(f\sigma_{8})'+\Big(\frac{\mathcal{H}'}{\mathcal{H}}+1\Big)f\sigma_{8}+\frac{\sigma_{8}^{0}}
{\delta_{_{\!\!\!\!\sim}}(0,k)}\frac{k^{2}}{\mathcal{H}^{2}}\Psi=0\,,
\end{equation}
where the Newtonian potential $\Psi$ from (\ref{pwe}) is proportional to $G_{\rm{eff}}$ and of
opposite sign. A recent negative $G_{\rm{eff}}$ gives a positive $\Psi$, as already known, and decreases $f\sigma_{8}$. A similar behavior in $G_{\rm{eff}}$ was observed in a specific nonlocal model of modified gravity that also led to a prediction of less growth than $\Lambda$CDM \cite{henrik}.

\begin{figure}[t!]
\begin{minipage}{.5\linewidth}
\centering
\subfloat[]{\label{main:a}\includegraphics[scale=.41]{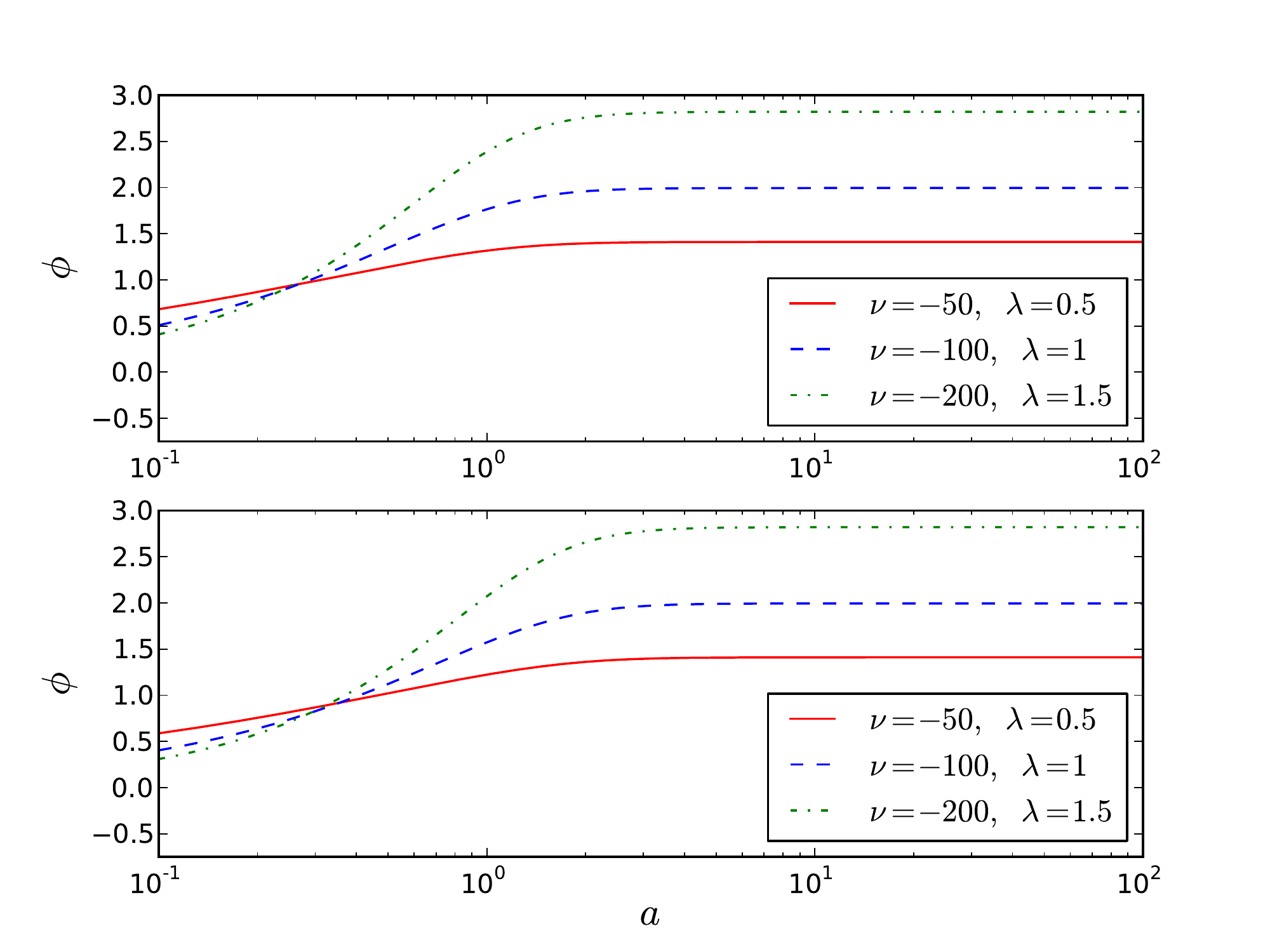}}
\end{minipage}%
\begin{minipage}{.5\linewidth}
\centering
\subfloat[]{\label{main:b}\includegraphics[scale=.41]{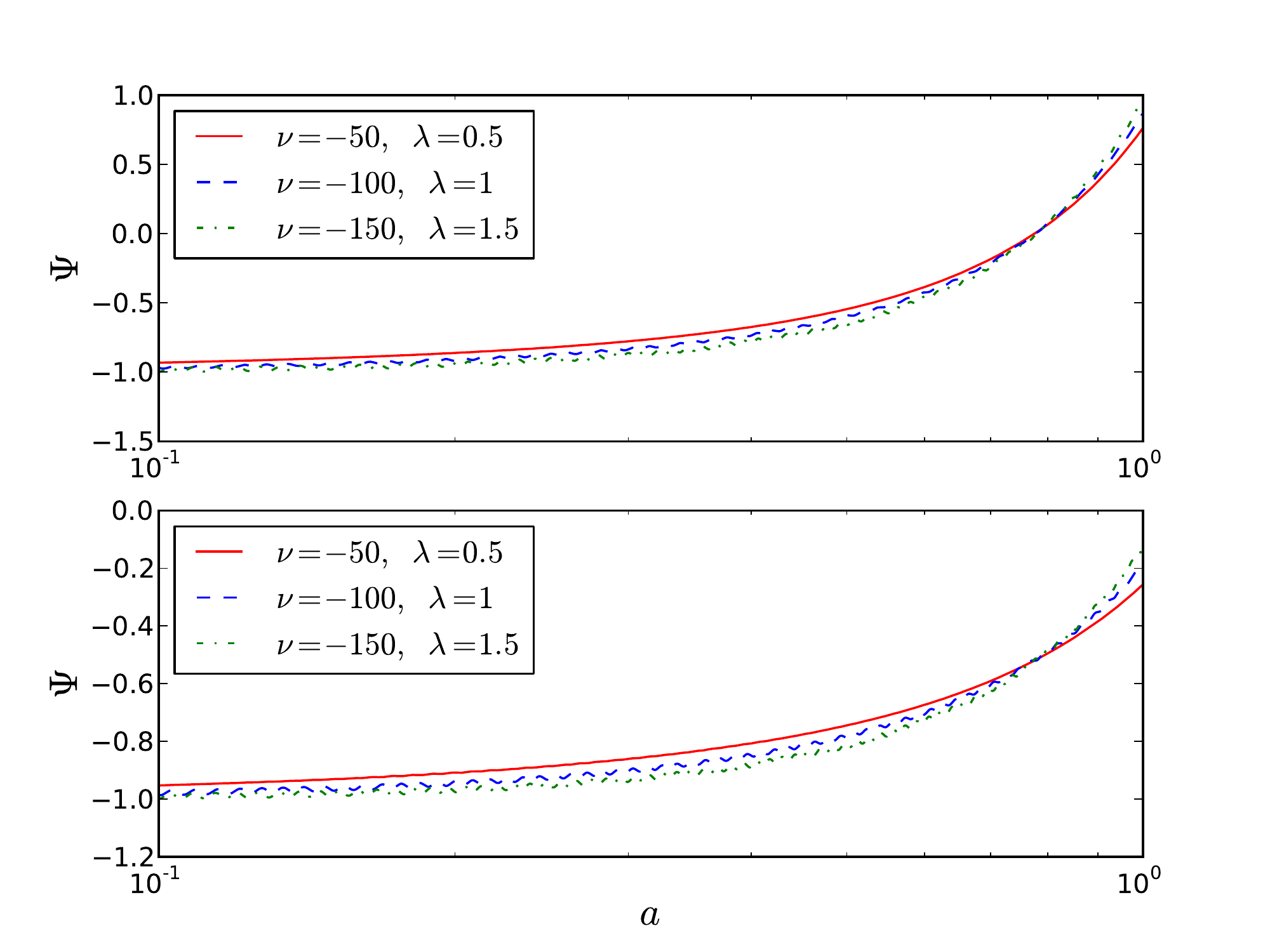}}
\end{minipage}
\caption{\label{psi_phi}We plot the evolution of the scalar field $\phi$ (background part only) and
the Newtonian potential $\Psi$ as a function of scale factor, for the CBD model. On the top plots, we
have $\Omega_{\rm{DE}}^{0}\approx 0.7$, while on the bottom plots we require
$\Omega_{\rm{DE}}^{0}\approx 0.5$.}
\end{figure}

We have tested numerically that, as long as one requires $\Omega_{\rm{DE}}^{0} \approx 0.7$, the
above change in sign of $G_{\rm{eff}}$ or $\Psi$ close to the present persists, independently of
the parameters or the initial conditions of the background evolution, even set at different redshifts.
We have actually found very special values of the parameters (with $\lambda<0$) and initial conditions,
consistent with $\Omega_{\rm{DE}}^{0} \approx 0.7$, such that $G_{\rm{eff}}^{0}>0$, however, the whole
cosmology arising is physically unacceptable. Also, $G_{\rm{eff}}$ can remain permanently positive if
the requirement of background viability is relaxed and $\Omega_{\rm{DE}}^{0}$ is set to a value of
approximately $1/2$. We can not, for now, provide a definite proof on the inevitability of the change
of sign in $G_{\rm{eff}}$ for reasonable evolutions of the CBD theory, however it seems that
the scalar field evolves toward $\varphi_{\infty}$ as we progress into the far future, and
before reaching the present-time, will already have crossed the critical value $\varphi_{\rm{c}}$.
In Fig.~\ref{psi_phi}, we plot the evolution of the scalar field $\phi$ (background part only) as a
function of the scale factor $a$ (which we extend beyond $a=1$) for different parameters $\lambda$
and $\nu$, together with the evolution of $\Psi$. As we can see, the crossing in $\Psi$ is inevitable,
unless one relaxes the requirement of having $\Omega_{\rm{DE}} \approx 0.7$ today. Only when we take
$\Omega_{\rm{DE}}^{0} \approx 0.5$ is the crossing in $\Psi$  not verified, as the scalar field tends to
same asymptotic value more slowly.

Negative values of $G_{\rm{eff}}$ is a fundamental issue and may jeopardize
the viability of the theory on the smallest scales, however, this does not mean that the CBD theory
should be ruled out immediately. First, it is possible that the gravitational constant that controls
the gravitational effects in a static spherically symmetric configuration around a central mass
is unrelated to the above $G_{\rm{eff}}$. This issue can be resolved if local solutions are
found and a PPN analysis is performed. There could also be a screening mechanism ensuring the
suppression of the additional interaction mediated by the theory on the smallest scales and, hence,
hiding any evidence of this change of sign in $G_{\rm{eff}}$. Another option could be that
the theory does not couple to baryons, and hence only impact dark matter, allowing it to modify
galactic dynamics without affecting ordinary matter and passing laboratory tests of gravity.
Lastly, there is the possibility of considering a potential $V(\phi)$ that does not have
to be dominant today, but could provide the necessary contribution to the dark energy density that
would prevent the scalar field crossing the critical value that changes the sign of $G_{\rm{eff}}$.
It would also be interesting to perform a complete dynamical analysis of the equations of motion of
the theory, since this would allow to make a more definitive statement on the behavior of
$G_{\rm{eff}}$ and, eventually, find background attractor solutions that could avoid this problem
altogether.

\section{Conclusions \label{conclusion}}

In this work, we focused on one of three generalizations of the standard Brans-Dicke gravity (BD),
named as complete Brans-Dicke theories (CBD) \cite{Kofinas:2015nwa}. These were derived at the level of
the field equations by analyzing exhaustively the Bianchi identities, while maintaining the BD scalar
field wave equation and relaxing the standard matter conservation.

For this particular model, which for brevity we also refer to as CBD, there is one new parameter $\nu$
that mediates the interaction between the dark sectors. It had been previously shown that, for negative
values of this parameter, the theory was able to produce accelerating cosmological solutions today
without the presence of a potential $V(\phi)$ \cite{Kofinas:2016fcp}. Here, we have extended the
applicability of these solutions to high redshifts, which, as we show in Sec.~\ref{seccosmo}, yield
a stable matter domination regime that is gradually overtaken by the dark energy component to yield
acceleration today. Moreover, we obtain a nice fit to the low-redshift supernovae data.
For our background solutions, we assume slow-roll initial conditions, with the
initial value of the scalar field being found by requiring $\Omega_{\rm{DE}}\approx 0.7$ today.

We then study the evolution of linear perturbations in the CBD theory in order to understand the impact
it can have on the large scale structure of the Universe we observe. We present the full set of
perturbed gravitational equations in both the Newtonian and synchronous gauges. One feature that becomes
immediately obvious, and is transversal to most modified gravity theories, is the dynamical anisotropy
between the gravitational potentials, dependent on the perturbation $\chi$ of the scalar field.

In particular, $\chi$ evolves according to a damped harmonic oscillator subjected to an external force
proportional to the matter perturbation $\delta_{_{\!\!\!\!\sim}}$. At late-times, as we show in
Sec.~\ref{lensingpot}, after the oscillations have been damped out, $\chi$ is pushed toward larger
values relatively to its equilibrium initial position which we set to zero. In turn, this is manifested
in the lensing potential $\Phi_{+}$ which exhibits an unusual growth at late-times. Hence,
$\Phi_{+}$ not only resists the expanding background, but does increasing in amplitude, in a clear
departure from $\Lambda$CDM, where the perturbations are expected to decay once $\Lambda$ starts to
dominate. This behavior becomes clear looking at the sub-horizon quasi-static approximation for the
evolution of the Newtonian potentials, which we present in Sec.~\ref{subhorsec}. Then, the lensing
potential is directly proportional to $\chi$, and hence follows its late-time behavior, growing as
we approach $a=1$.

Another interesting feature is that the evolution of all perturbations in the CBD theory is
scale-dependent at early times. At late-times the gravitational potentials and the
scalar field perturbation become scale-independent, what can be explained through the
sub-horizon approximation and also be observable in our numerical results. This is verified in
non-massive standard Brans-Dicke gravity as well \cite{defelice}.

We have also studied the evolution of the growth rate for the complete Brans-Dicke theory. We have
concluded that the CBD theory predicts less growth than $\Lambda$CDM, and could produce a better fit
to existent $f\sigma_{8}(z)$ data from RSD observations. This fact is clearly explained in the
sub-horizon approximation, where the behaviour of $f\sigma_{8}(z)$ is controlled by the time-time
Newtonian potential $\Psi$. Contrary to the behavior of the lensing potential, $\Psi$ passes
from negative to positive values recently, and hence, as the scalar field grows in time, the
theory predicts less growth than $\Lambda$CDM. However, in parallel with the sign change of $\Psi$
in the sub-horizon scales, the effective gravitational constant $G_{\rm{eff}}$ for the
perturbations also changes sign and becomes negative recently. This effect seems to persist
independently of the choice of the parameters or the initial conditions of any reasonable background
evolution, and may jeopardize the validity of the theory on the smallest scales. It is
premature to decide on this before local spherically symmetric solutions are found
and the existence of screening mechanisms is investigated that might suppress the additional
interaction mediated by the theory in order to pass the stringent solar-system tests of gravity.
Other options would be the decoupling of baryons from the theory which would alleviate the small
scales constraints on the theory, or the existence of a potential preventing the sign change of
$G_{\rm{eff}}$. Finally, the study of the background attractor solutions through a dynamical system
analysis should allow a more decisive statement on the behavior of $G_{\rm{eff}}$.

\section*{Acknowledgements}

The authors would like to thank Emmanuel Saridakis for helpful discussions about the background solutions of the complete Brans-Dicke theory. We also thank Luca Amendola and Valeria Pettorino for helpful discussions and comments. The work of N.A.L. is supported by the DFG through the Transregional Research Center TRR33 “The Dark Universe”.

\appendix
\section{Jordan frame perturbation equations}
\label{perturbations}

We present here some perturbed geometric quantities used for deriving the perturbed equations
of motion. As for the Christoffel symbols we have
\begin{widetext}
\begin{eqnarray}{\label{christoffel}}
&&\delta \Gamma_{\,\,\,00}^{0} = \dot{A} Y\,\,\,, \hspace{0.25 cm}  \delta \Gamma_{\,\,\,0i}^{0}
 = -\PC{kA +\mathcal{H}B}Y_{i} \\
&&\delta \Gamma_{\,\,\,ij}^{0} = \PC{-2 \mathcal{H}A + \frac{k}{3}B + 2\mathcal{H}H_L +
 \dot{H}_L} \delta_{ij} Y + \PC{-kB + 2\mathcal{H} H_T + \dot{H}_T} Y_{ij} \\
&&\delta \Gamma_{\,\,\,00}^{i} = -\PC{kA + \dot{B} + \mathcal{H}B} Y^{i} \\
&&\delta \Gamma_{\,\,\,0j}^{i} = \dot{H}_L \delta^{i}_{j} Y + \dot{H}_T Y^{i}_{\,\,\,j} \\
&&\delta \Gamma_{\,\,jk}^{i} = -k H_L \PC{\delta^{i}_{j} Y_k + \delta^{i}_{k} Y_j -
 \delta_{jk}Y^{i}} + \mathcal{H}B \delta_{jk} Y^{i} + H_T \PC{Y^{i}_{\,\,\,j,k}
 + Y^{i}_{\,\,\,k,j} - Y_{jk}^{\,\,\,\,,i}}\,.
\end{eqnarray}
\end{widetext}
Indices in $Y_{i}$, $Y_{ij}$ are raised with $\delta^{ij}$. The perturbed Ricci tensor and Ricci
scalar are
\begin{widetext}
\begin{eqnarray}
\label{riccisjordan}
&&\!\!\delta R = \frac{2}{a^2}\PR{-6 \frac{\ddot{a}}{a} A - 3\mathcal{H}\dot{A} + k^2A + k\dot{B}
+ 3k\mathcal{H}B + 9\mathcal{H}\dot{H}_L + 3\ddot{H}_L + 2k^2\PC{H_L + \frac{H_T}{3}}} Y \\
\label{riccitensor}
&&\!\!\delta R_{00} = - \PR{k^2A - 3\mathcal{H}\dot{A} + k\PC{\dot{B} + \mathcal{H}B} + 3\ddot{H}_L
+ 3\mathcal{H}\dot{H}_L}Y\\
&&\!\!\delta R_{0i} = \PR{-\PC{\frac{\ddot{a}}{a} + \mathcal{H}^2}B - 2k\mathcal{H}A + 2k\dot{H}_L
+ \frac{2}{3}k\dot{H}_T}Y_i \\
&&\!\!\delta R_{ij} = \Bigg[ -2\PC{\frac{\ddot{a}}{a} + \mathcal{H}^2}A - \mathcal{H}\dot{A}
+ \frac{k^2}{3}A + \frac{k}{3}\PC{\dot{B} + 5\mathcal{H}B} + \ddot{H}_L + 5\mathcal{H}\dot{H}_L
+ 2\PC{\frac{\ddot{a}}{a} + \mathcal{H}^2}H_L + \frac{4k^2}{3}\PC{H_L + \frac{H_T}{3}} \Bigg]
\delta_{ij}Y \nonumber \\
&&\,\,\,\,\,\,\,\,\,\,\,\,\,\,\,
+\PR{-k^2A - k\PC{\dot{B} + \mathcal{H}B} + \ddot{H}_T + \mathcal{H}\dot{H}_T
+ 2\PC{\frac{\ddot{a}}{a}+\mathcal{H}^2}H_T - k^2\PC{H_L + \frac{H_T}{3}} + \mathcal{H}\PC{\dot{H}_T
- kB}}Y_{ij}\,.
\end{eqnarray}
\end{widetext}
The perturbations of the scalar field derivatives, due to $\delta g_{\mu\nu}$ and $\delta\phi$,
are given by the expressions
\begin{widetext}
 \begin{eqnarray}{\label{covarpert}}
&&\delta \PC{\nabla_{\mu} \nabla_{\nu} \phi} = \nabla_{\mu}\nabla_{\nu} (\delta \phi)
- \delta \Gamma_{\,\,\,\mu \nu}^{\lambda} \partial_{\lambda} \varphi \\
&&\delta \PC{\nabla^{\mu} \nabla_{\nu} \phi} = \nabla^{\mu}\nabla_{\nu} (\delta \phi)
+ \delta g^{\mu \lambda} \nabla_{\lambda} \nabla_{\nu} \phi
- g^{\mu \lambda} \delta \Gamma^{\kappa}_{\,\,\,\lambda \nu} \partial_{\kappa} \varphi\,,
 \end{eqnarray}
\end{widetext}
where $\delta \PC{\nabla_{\mu} \nabla_{\nu} \phi}=\nabla_{\mu} \nabla_{\nu} \phi-
\bar{\nabla}_{\mu}\bar{\nabla}_{\nu}\varphi$, $\delta \PC{\nabla^{\mu} \nabla_{\nu} \phi}=
g^{\mu\lambda}\nabla_{\nu} \nabla_{\lambda} \phi
-\bar{g}^{\mu\lambda}\bar{\nabla}_{\nu}\bar{\nabla}_{\lambda}\varphi$, $\nabla$ denotes the covariant
derivative with respect to the perturbed metric $g_{\mu\nu}$, $\bar{\nabla}$ denotes the covariant
derivative with respect to the background metric $\bar{g}_{\mu\nu}$ and $\phi=\varphi+\delta\phi$.
Since $\delta\phi=\chi(t)Y$, we get
\begin{widetext}
\begin{eqnarray}
&&\delta \PC{\nabla_{0} \nabla_{i} \phi} = \PR{-k \dot{\chi}
 + k \mathcal{H} \chi + \dot{\varphi}\PC{kA + \mathcal{H}B}}Y_i \\
&&\delta \PC{\nabla_{0} \nabla_{0} \phi} = \PC{\ddot{\chi} - \mathcal{H} \dot{\chi}
- \dot{\varphi} \dot{A}} Y\\
&&\delta \PC{\nabla_{i} \nabla_{j} \phi} = \PR{- \mathcal{H} \dot{\chi} -\frac{k^2}{3}\chi
 + \dot{\varphi}\PC{2\mathcal{H}A-\frac{k}{3}B-2\mathcal{H}H_L-\dot{H}_L}}
 \delta_{ij}Y
 \nonumber \\
&&\hspace{2.0 cm} + \PR{k^2 \chi + \dot{\varphi}\PC{kB
 - 2 \mathcal{H} H_T - \dot{H}_T}}Y_{ij} \\
&&\delta \PC{\nabla^{0} \nabla_{0} \phi} = \frac{1}{a^2}\PC{-\ddot{\chi}
 + \mathcal{H} \dot{\chi} + 2 \ddot{\varphi}A - 2\dot{\varphi}\mathcal{H}A + \dot{\varphi} \dot{A}}Y \\
&&\delta \PC{\nabla^{0} \nabla_{i} \phi} = \frac{1}{a^2}\PC{k \dot{\chi}
 - k\mathcal{H} \chi - k\dot{\varphi}A}Y_i \\
&&\delta \PC{\nabla^{i} \nabla_{j} \phi} = \frac{1}{a^2}\PR{k^2 \chi
 + \dot{\varphi} \PC{kB - \dot{H}_T}}Y^{i}_{\,\,\,j} + \frac{1}{a^2}\PR{-\mathcal{H}\dot{\chi}
 -\frac{k^2}{3}\chi+\dot{\varphi}\PC{2\mathcal{H}A-\frac{k}{3}B-\dot{H}_L}}\delta^{i}_{j}Y\\
&&\delta \PC{\nabla^{i} \nabla_{i} \phi} = \frac{1}{a^2}\PR{- 3\mathcal{H} \dot{\chi}
-k^2 \chi + \dot{\varphi}\PC{6 \mathcal{H}A - kB - 3\dot{H}_L}}Y
\end{eqnarray}
\end{widetext}

\bibliography{cbd_pert_bib}

\end{document}